%% file: main.tex
\documentclass[
  journal=pasa,
  manuscript=research-paper,
  year=2021,
  volume=37,
]{cup-journal}

\usepackage[utf8]{inputenc}
\usepackage[T1]{fontenc}
\usepackage{xcolor}
\usepackage{adjustbox}
\usepackage{microtype,siunitx}
\usepackage[backref=page]{hyperref}
\definecolor{dodgerblue}{RGB}{30, 144, 255}
\definecolor{crimson}{RGB}{220, 20, 60}
\definecolor{darkerblue}{RGB}{0, 0, 139}
\definecolor{darkred}{RGB}{150,20,20}
\definecolor{cpink}{RGB}{243, 141, 252}

\hypersetup{colorlinks,citecolor=dodgerblue,linkcolor=blue,urlcolor=magenta}
\usepackage{enumitem, xcolor}
\let\svitem\item

\usepackage{booktabs} 
\usepackage{threeparttable,longtable}  
\usepackage{makecell}

\usepackage[skip=0.5ex]{subcaption}
\usepackage{cprotect}
\renewcommand\thesubfigure{(\roman{subfigure})}

\usepackage{multirow}

\title{The Rapid ASKAP Continuum Survey (RACS) VI: The RACS-high 1\,655.5\,MHz images and catalogue}

\author{S.~W.~Duchesne}
\affiliation{CSIRO Space and Astronomy, PO Box 1130, Bentley WA 6102, Australia}
\email[S.~W.~Duchesne]{Stefan.Duchesne@csiro.au}

\author{K.~Ross}
\affiliation{International Centre for Radio Astronomy Research, Curtin University, Bentley, WA 6102, Australia}

\author{A.~J.~M.~Thomson}
\affiliation{CSIRO Space and Astronomy, PO Box 1130, Bentley WA 6102, Australia}

\author{E.~Lenc}
\affiliation{CSIRO Space and Astronomy, PO Box 76, Epping, NSW, 1710, Australia}

\author{Tara~Murphy}
\affiliation{Sydney Institute for Astronomy, School of Physics, University of Sydney, NSW 2006, Australia}

\author{T.~J.~Galvin}
\affiliation{CSIRO Space and Astronomy, PO Box 1130, Bentley WA 6102, Australia}

\author{A.~W.~Hotan}
\affiliation{CSIRO Space and Astronomy, PO Box 1130, Bentley WA 6102, Australia}

\author{V.~A.~Moss}
\affiliation{CSIRO Space and Astronomy, PO Box 76, Epping, NSW, 1710, Australia}
\alsoaffiliation{Sydney Institute for Astronomy, School of Physics, University of Sydney, NSW 2006, Australia}

\author{Matthew~T.~Whiting}
\affiliation{CSIRO Space and Astronomy, PO Box 76, Epping, NSW, 1710, Australia}

\doi{10.1017/pasa.20XX.XX}

\received {dd Mmm YYYY}
\revised  {dd Mmm YYYY}
\accepted {dd Mmm YYYY}
\published{22 September 2020}


\definecolor{dodgerblue}{RGB}{30, 144, 255}

\newcommand{\corrs}[1]{{#1}}

\begin{document}

\input{content}


\bibliography{bib}

\appendix

\input{appendices}

\end{document}

%% file: content.tex
\begin{abstract}
We have conducted a widefield, wideband, snapshot survey using the Australian SKA Pathfinder (ASKAP) referred to as the Rapid ASKAP Continuum Survey (RACS). RACS covers $\approx 90$\,\% of the sky, with multiple observing epochs in three frequency bands sampling the ASKAP frequency range of 700--1\,800\,MHz. This paper describes the third major epoch at \corrs{1\,655.5}\,MHz, RACS-high, and the subsequent imaging and catalogue data release. The RACS-high observations at \corrs{1\,655.5}\,MHz are otherwise similar to the previously released RACS-mid (at 1367.5\,MHz), and were calibrated and imaged with minimal changes. From the 1\,493 images covering the sky up to declination $\approx +48\degr$, we present a catalogue of 2\,677\,509 radio sources. The catalogue is constructed from images with a median root-mean-square noise of $\approx 195$\,\textmu Jy\,PSF$^{-1}$ (point-spread function) and a median angular resolution of $11\farcs8 \times 8\farcs1$. The overall reliability of the catalogue is estimated to be 99.18\,\%, and we find a decrease in reliability as angular resolution \corrs{improves}. We estimate the brightness scale to be accurate to 10\,\%, and the astrometric accuracy to be within $\approx 0\farcs6$ in right ascension and $\approx 0\farcs7$ in declination after correction of a systematic declination-dependent offset. All data products from RACS-high, including calibrated visibility datasets, images from individual observations, full-sensitivity mosaics, and the all-sky catalogue are available at the CSIRO ASKAP Science Data Archive. 

\end{abstract}

\section{Introduction}
\label{sec:int}

\defcitealias{racs1}{Paper~I}
\defcitealias{racs2}{Paper~II}
\defcitealias{Thomson2023}{Paper~III}
\defcitealias{racs-mid}{Paper~IV}
\defcitealias{racs-mid2}{Paper~V}

The Rapid ASKAP Continuum Survey \footnote{\url{https://research.csiro.au/racs/}.} \citep[RACS;][]{racs1} is an ongoing widefield radio survey being conducted by the Australian SKA Pathfinder \citep[ASKAP;][]{Hotan2021}, operated by CSIRO \footnote{Commonwealth Scientific and Industrial Research Organisation.} on Inyarrimanha Ilgari Bundara, the CSIRO Murchison Radio-astronomy Observatory. RACS is designed as a snapshot survey in multiple frequency bands across the 700--1\,800\,MHz available to ASKAP, covering $\approx 90$\% of the sky in approximately two weeks of total integration time for each frequency band. The survey speed of ASKAP is partially enabled by the phased array feeds (PAFs) on each of its 36 antennas. The PAFs provide 36 primary beams that are arranged into overlapping footprints covering a $\approx 31$\,deg$^{2}$ field of view (FoV) per observation.   

RACS has had two major data releases so far, one in the lowest part of ASKAP's observing band at 887.5\,MHz \citep[RACS-low;][hereinafter \citetalias{racs1}]{racs1} and one in the middle part of the band at 1\,367.5\,MHz \citep[RACS-mid;][hereinafter \citetalias{racs-mid}]{racs-mid}. In addition to the main data releases, comprising images and calibrated visibilities, all-sky catalogues have also been produced for both RACS-low \citep[][hereinafter \citetalias{racs2}]{racs2} and RACS-mid \citep[][hereinafter \citetalias{racs-mid2}]{racs-mid2}. RACS data products are made available through the CSIRO ASKAP Science Data Archive\footnote{\url{https://data.csiro.au/domain/casda}.} \citep[CASDA;][]{casda,Huynh2020} as part of project AS110. The third epoch of RACS is in the high band at 1\,655.5\,MHz (RACS-high) and was observed over 2021--2022. Following RACS-high, two further epochs were completed in the low band at 887.5\,MHz and 943.5\,MHz and will be discussed elsewhere.

The initial data releases for RACS epochs focus on total intensity (Stokes $I$) and circular polarization (Stokes $V$) images and catalogues, though RACS observations contain all four polarization products, allowing all Stokes parameters ($I$, $V$, and linear polarization in $Q$ and $U$) to be imaged. To leverage the linear polarization component of RACS, \citet[][hereinafter, \citetalias{Thomson2023}]{Thomson2023} introduced Spectra and Polarisation In Cutouts of Extragalactic Sources from RACS (SPICE-RACS). The first data release from SPICE-RACS made use of RACS-low data covering a $\approx 1\,300$-deg$^2$ region as a pilot to the all-sky project. An initial all-sky catalogue of RACS-low3 has also been created and is being used by Thomson et al. (in preparation) to guide linear polarization processing of the RACS-low3 component of SPICE-RACS. Those data will be described in a future paper.  

 RACS has already been used for validation and calibration of datasets from ASKAP, including primary beam modelling of archival datasets \citep{Duchesne2024} and as a quality assurance cross-match for some of the main ASKAP surveys, including the First Large Absorption Survey in HI \citep[FLASH;][]{Yoon2024} and the Evolutionary Map of the Universe \citep[EMU;][]{Norris2021} survey (Hopkins et al., submitted), with RACS being used in analyses of data products from other instruments \citep[e.g.][]{Frail2024,Rajwade2024}. While the main goal of RACS is to generate a sky model for calibration of future ASKAP observations, the individual bands themselves (low, mid, and high) provide some of the most sensitive and highest-resolution `all-sky' surveys performed at these frequencies to date. RACS provides valuable information for wideband spectral modelling \citep[e.g.][]{Kerrison2024,Sanderson2024}, and its angular resolution and sensitivity provides the means to detect and characterise high redshift radio galaxies and active galactic nuclei (AGN) \citep[e.g.][]{Broderick2022,Broderick2024} and distant, bright quasi-stellar objects \citep{Ighina2024a,Ighina2024b}. \corrs{With such wide area sky coverage, the RACS catalogues have also seen use in investigating the cosmic radio dipole \citep[e.g.][]{Wagenveld2023,Oayda2024}.} Looking within our own Galaxy, RACS data across the band have been used for searches for radio emission from stars in both total intensity and circular polarization \citep{Rose2023,Driessen2023,Driessen2024,Pritchard2024}, radio counterparts of novae \citep{Gulati2023}, and for characterising other Galactic radio sources \citep[e.g.][]{vandenEijnden2022,vandenEijnden2024}.

{As RACS provides additional epoch-specific information, the images and catalogues have been used extensively in time-domain astronomy, and is included as additional epochs for the Variability And Slow Transients project \cite[VAST;][]{Murphy2021}. For example, data products from RACS have enabled detection of radio emission from tidal disruption events \citep{Anumarlapudi2024,Dykaar2024}, radio variability in stars \citep[e.g.][]{Pritchard2024}, and for measurements of variability in AGN \citep[e.g.][]{Ross2023}.}

{As a full epoch can be completed with approximately two weeks worth of observing time using relatively short snapshot observations, RACS does not put a significant strain on the ASKAP observing time available to its main surveys. As with other ASKAP surveys, the Commensal Real-time ASKAP Fast Transient (CRAFT) project has resulted in detection of fast radio bursts while observing various RACS epochs \citep{Shannon2024}.}

This paper discusses the data processing, data release, and all-sky source catalogue for RACS-high, providing a general overview of the data available. A summary of the currently-observed RACS epochs is provided in Table~\ref{tab:racs}.

\begin{table*}[t]
    \centering
    \begin{threeparttable}
    \caption{\label{tab:racs}RACS epochs currently observed and total intensity properties.}
    \begin{tabular}{c c c c c c c}\toprule
         Epoch name & Frequency & Date range \tnote{a}  & Median $\sigma_\text{rms}$ \tnote{b} & Median PSF \tnote{b} & $N_\text{sources}$ \tnote{c} & References \\
         & (MHz) &  & (\textmu Jy\,PSF$^{-1}$) & ($^{\prime\prime} \times ^{\prime\prime})$ & &\\[0.4em]\midrule
         RACS-low & 887.5 & 2019-04-21 to 2020-06-21 & 266 & $18.4 \times 11.6$ & 2\,297\,596 & \citet{racs1,racs2} \\
         RACS-mid & 1367.5 & 2020-12-20 to 2022-06-11 & 198 & $10.1 \times 8.1$ &  3\,105\,668 & \citet{racs-mid,racs-mid2}\\
         RACS-high & 1655.5 & 2021-12-24 to 2022-03-03 & 209 & $8.6 \times 6.3$ & 2\,677\,509 & This work \\
         RACS-low2 & 887.5 & 2022-03-18 to 2022-06-21 & - & - & - & - \\
         RACS-low3 & 943.5 & 2023-12-20 to 2024-04-06 & 205 & $13.4 \times 11.0$ & - & - \\[0.4em]\bottomrule
\end{tabular}
\begin{tablenotes}[flushleft]
{\footnotesize \item[a] {Overall date range for the entire epoch, including re-observations usually taken at a later date.} \item[b] Based on the single-observation images. \item[c] From the primary all-sky catalogue, if available.}

\end{tablenotes}
\end{threeparttable}
\end{table*}

\section{Data}

RACS-high is observed in an identical fashion to RACS-mid, using the same \texttt{closepack36} footprint for the PAF (see figure~1 in \citetalias{racs-mid}) and the same tiling of the celestial sphere \citepalias[see figure~3 in][]{racs-mid}. This results in the same coverage at the cost of inhomogeneous sensitivity across the full FoV due to the smaller primary beam (see Section~\ref{sec:tile_sensitivity}). While some of the high band data are also flagged, we still retain a full 288\,MHz bandwidth dataset for RACS-high. RACS pointing directions are referred to as `fields' and from RACS-mid onwards labelled by truncating their central $(\alpha_\text{J2000},\delta_\text{J2000})$ coordinates: \texttt{RACS\_HHMM$\pm$DD}. Individual observations for ASKAP are referred to as `Scheduling Blocks' with each observation referred to by a unique Scheduling Block ID (SBID) \footnote{As an example, the observations are usually referred to in the form \corrs{`SB34668'}}. For RACS-mid onwards, all observations contain a single field. After initial processing of each SBID, the images are rapidly inspected for obvious issues that would require a re-observation. 91 fields were re-observed and so have duplicate SBIDs available. In all cases, both observations are usable for science.   

\subsection{Primary beam models}\label{sec:holography}

\begin{table}[t]
    \centering
    \begin{threeparttable}
    \caption{\label{tab:holography}Holography observation details and associated science SBIDs.}
    \begin{tabular}{ c c c}\toprule
        Holography SBID & Science SBID range \tnote{a} & Date range for science SBIDs \tnote{a} \\\midrule
        34422 & 34668--35367 & 2021-12-24 to 2022-01-05  \\
        35847 & 35397--37101 & 2022-01-06 to 2022-02-03 \\
        37170 & 37145--37816 & 2022-02-04 to 2022-03-03 \\\bottomrule
    \end{tabular}
    \begin{tablenotes}[flushleft]
    {\footnotesize \item[a] Inclusive.}
    \end{tablenotes}
    \end{threeparttable}
\end{table}

For RACS-high, we also made holographic observations using PKS~B0407$-$658 to map the frequency-dependent primary beam model for each of the PAF beams. Over the course of processing RACS-mid \citepalias{racs-mid} and SPICE-RACS DR1 \citepalias{Thomson2023}, the primary beam shapes and positions were found to change significantly enough when beamforming weights were updated. These changes can result in significant brightness scale errors (or widefield leakage in polarization data products) if uncorrected or unmodelled. To account for this, the Observatory now performs holography observations whenever there are major updates to the beamformer weights, and generates the corresponding primary beam models. This was introduced for RACS-high and is present for all subsequent RACS epochs. For RACS-high we have three separate primary beam models that are applied across the survey, summarised in Table~\ref{tab:holography}. 

While the Stokes $I$ total intensity models are used, due to an error during processing of the holography the widefield leakage models of Stokes $I$ into $V$ were not complete. The raw holography observational data has since been deleted. Because of this we opt to forgo application of widefield leakage corrections for the RACS-high Stokes $V$ data products.

\subsection{Calibration and imaging}\label{sec:imaging}

\begin{figure}[t]
    \centering
    \includegraphics[width=1\linewidth]{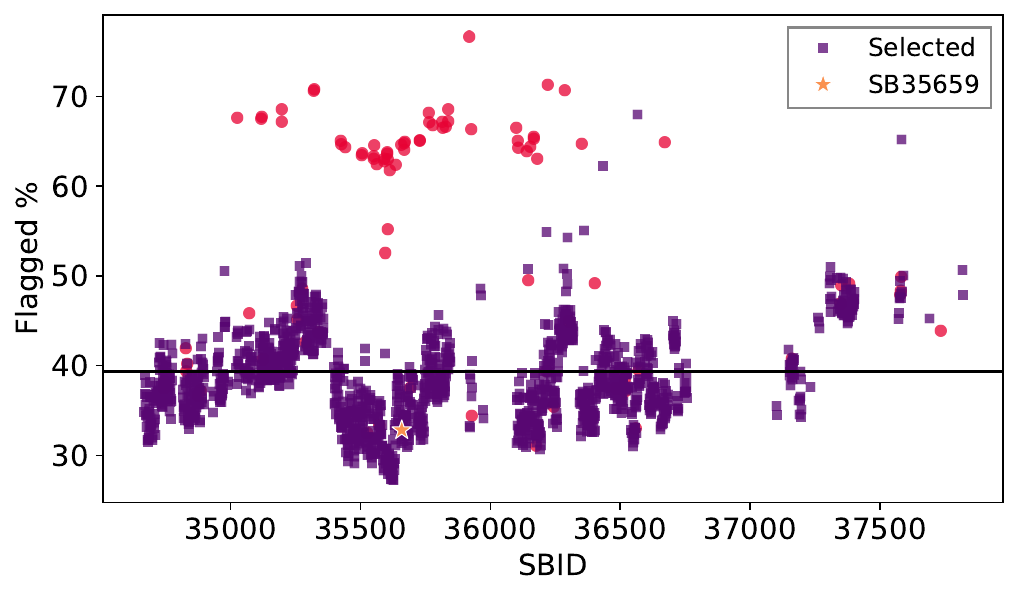}
    \caption{\label{fig:flag_summary} \corrs{Mean percentage of flagged data averaged across all PAF beams for each SBID. The SBIDs selected for mosaicking (Section~\ref{sec:selection}) are highlighted, and SB35659 that uses additional \texttt{CASA} flagging tasks is also shown. The mean percentage (39\%) is show as a horizontal black line.}}
\end{figure}

Calibration and imaging is performed with the \texttt{ASKAPsoft} pipeline (version 2.9.11) on the Pawsey Supercomputing \corrs{Research} Centre cluster \textit{Setonix} through the Pawsey Partner Allocation scheme under project \texttt{pawsey1014}. Imaging and calibration is similar to RACS-mid with only a few exceptions. The first exception is that \textit{all} images are created after excluding baselines of $<100$\,m, as opposed to a manual selection of SBIDs to apply this cut to. We use this cut to reduce large-scale ripples from off-axis extended sources, including interference from the Sun during a significant subset of observations. The baseline cut also helps to control artefacts originating from poorly modelled extended emission, particularly in the Galactic Plane. We find this baseline cut does not appreciably reduce sensitivity for comparatively quiet fields (a few \textmu Jy\,PSF$^{-1}$ difference). 

\corrs{Figure~\ref{fig:flag_summary} shows the mean percentage of flagged data, averaged over all PAF beams for each SBID. The overall flagged data percentage is ($39 \pm 7$)\%, ranging from 27--77\% across the survey. Note that while default pipeline settings for RFI flagging were sufficient for almost all of the observations, beam 32 of SB35659 required an additional round of automated flagging via the \texttt{rflag} algorithm implemented in CASA\footnote{Common Astronomy Software Applications, \citep{casa,casa2022}}. SB35659 is highlighted for reference on Figure~\ref{fig:flag_summary}, which has a below average percentage of data flagged overall. Beam 32 from SB35659, however, has 49\% of the data flagged whereas other beams average 32\%. On Figure~\ref{fig:flag_summary} we also highlight SBIDs selected for mosaicking, discussed in Section~\ref{sec:selection}.}

{Over the course of imaging for RACS-mid, we identified fields with significant artefacts from off-axis bright sources. These sources were peeled or subtracted from the affected datasets (see section~2.3.3 in \citetalias{racs-mid}), and this process is repeated for RACS-high with some changes. We updated the list of sources to peel to include a total 81 sources across 167 SBIDs. The implementation of peeling for RACS is made with a bespoke script inserted prior to self-calibration, and uses a new mode in the peeling pipeline, \texttt{PotatoPeel}\footnote{`Peel out that annoying, terrible object': \url{https://gitlab.com/Sunmish/potato}.}. The new mode allows multiple directions (i.e.\,sources) to be peeled successively in order of apparent brightness, as in traditional peeling \citep{Noordam2004,Smirnov2011b}.}

The self-calibration and imaging process for RACS-high is identical to RACS-mid (see section 2.3 in \citetalias{racs-mid2} for detail), with no significant \texttt{ASKAPsoft} pipeline changes in the interim. Creation of the linear mosaics and subsequent source-finding is similar to RACS-mid, with the exception of the primary beam models discussed above and the lack of widefield leakage correction in the Stokes $V$ images. Images are trimmed to a least bounding box to remove extraneous NaN pixels at the edge of the PAF mosaic\footnote{Except for SB34732---due to pipeline dependency errors the Stokes $V$ images were not trimmed alongside the Stokes $I$ images and as a result have different image sizes, though still have the correct World Coordinate System header information individually. An automated check of the individual image dimensions was implemented after this incident.}. 

For quality assurance work, the source-finder \texttt{Selavy} \citep[][]{duchamp} is used to construct source lists from the Stokes $I$ continuum mosaic for each SBID. The output source list is cross-matched with a range of sky surveys. This includes RACS-low and RACS-mid, as well as the Sydney University Molonglo Sky Survey \citep[SUMSS;][]{bls99,mmb+03}, NRAO\footnote{National Radio Astronomy Observatory.} VLA\footnote{Very Large Array.} Sky Survey \citep[NVSS;][]{ccg+98}, the VLA Sky Survey \citep[][]{vlass}, TIFR\footnote{Tata Institute for Fundamental Research.} GMRT\footnote{Giant Metrewave Radio Telescope.} Sky Survey \citep[TGSS alternate data release 1;][]{ijmf16}, and the International Celestial Reference Frame \citep[ICRF3;][]{Charlot2020}. \corrs{For the purpose of validation, the cross-matching is only performed for RACS-high sources with no neighbours within 90\,arcsec (i.e.\ isolated) and with a ratio of total flux density ($S_\text{total}$) to peak flux density ($S_\text{peak}$) satisfying $0.8 < S_\text{total}/S_\text{peak} < 1.2$ (i.e.\ compact).} These cross-matched data products per SBID are available through the RACS data repository\footnote{\url{https://bitbucket.csiro.au/projects/ASKAP_SURVEYS/repos/racs/browse}, with RACS-high under \texttt{epoch\_10}.}. We visually inspect the main Stokes $I$ and $V$ images for obvious errors and check basic summary statistics, including in the cross-matched source lists, to ensure no major issues are present. These visual inspections helped identify instances of poor flagging (noted above), and a handful of SBIDs that would benefit from a source being peeled. All imaging data products, except per-beam images, are then deposited onto CASDA. In addition to the main imaging data products, a collection of metadata (including calibration tables and primary beam models) is also included. In the following sections, we will describe the higher-order, non-\texttt{ASKAPsoft} pipeline data products, including updated images, which general users should consider using prior to the pipeline images and source lists. While not the main focus of this data release, the Stokes $V$ images are briefly assessed in Appendix~\ref{app:stokesv}.

\section{The all-sky catalogue and full-sensitivity images}

\subsection{Normalising the brightness scale}\label{sec:norm}

\begin{figure}[t]
    \centering
    \includegraphics[width=1\linewidth]{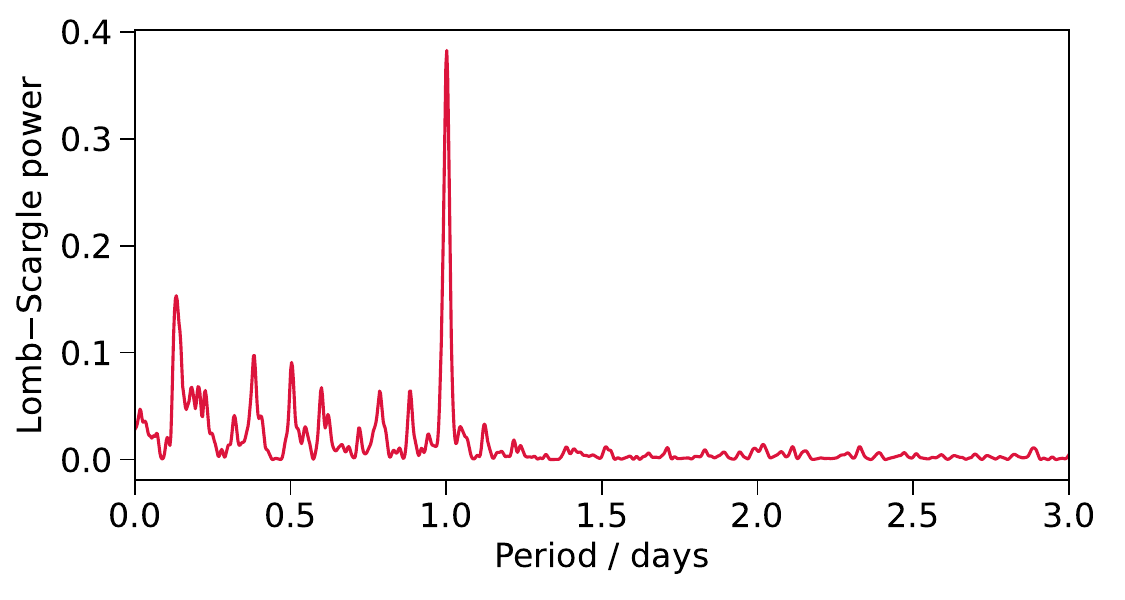}
    \caption{\label{fig:ls} The Lomb--Scargle periodogram of the median, observation-specific flux density ratios with respect to the NVSS.}
\end{figure}

\begin{figure}[t]
    \centering
    \includegraphics[width=1\linewidth]{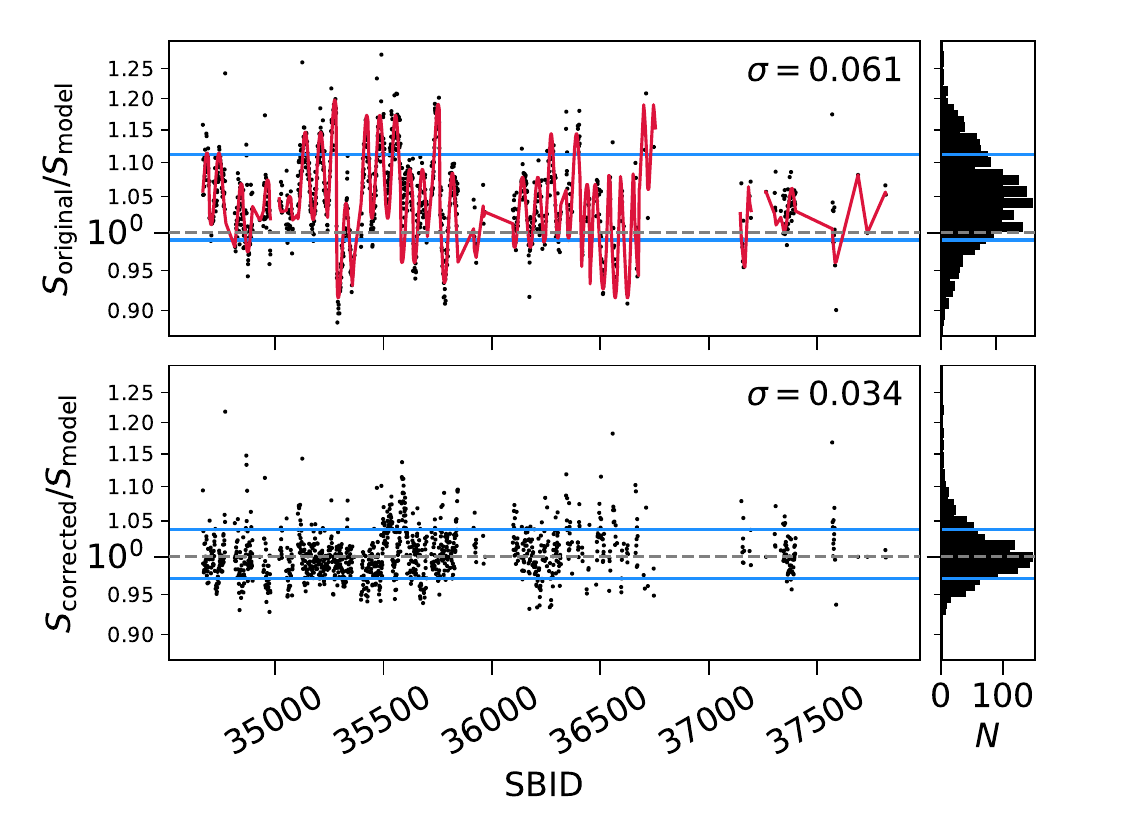}
    \caption{\label{fig:fluxscale} The median, observation-specific flux density ratios with respect to the NVSS, as a function of SBID. \emph{Top left panel.} The measured ratios (black points), with the scaled sinusoidal model overlaid (red line). \emph{Bottom left panel.} The ratios after application of the sinusoidal model. \emph{Right panels.} The histograms of the distributions. In all panels, the solid blue lines are drawn at $\pm 1\sigma$, centered on the mean. The dashed, gray lines correspond to ratios of 1.}
\end{figure}

\begin{figure*}[t]
    \centering
    \includegraphics[width=0.15\linewidth]{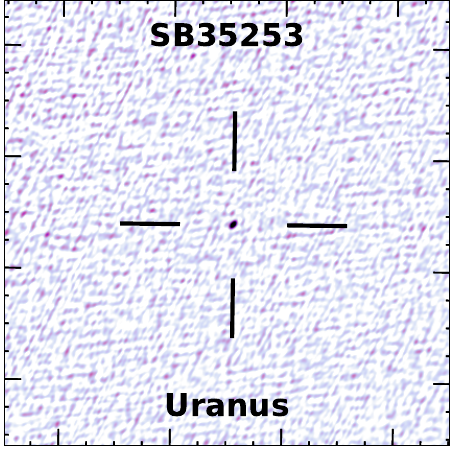}\,%
    \includegraphics[width=0.15\linewidth]{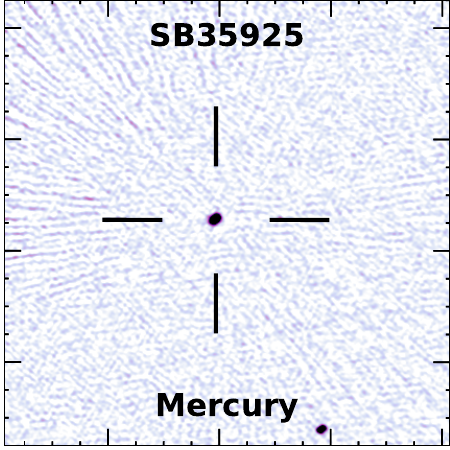}\,%
    \includegraphics[width=0.15\linewidth]{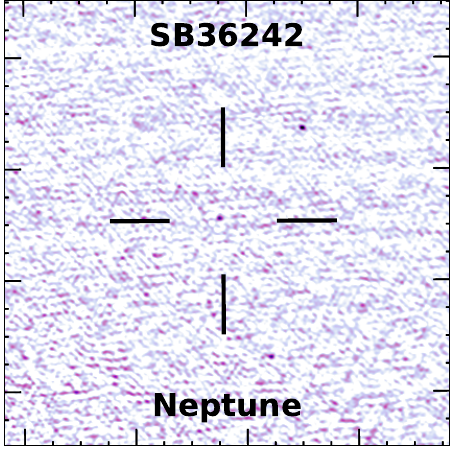}\,%
    \includegraphics[width=0.15\linewidth]{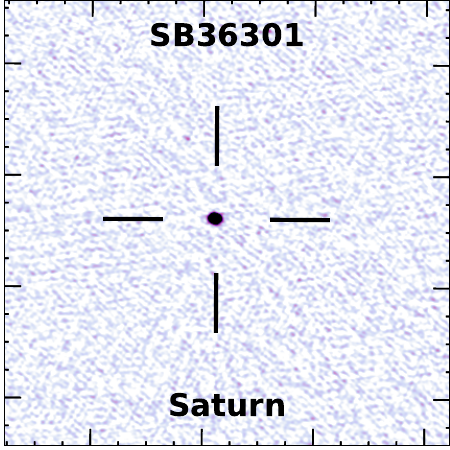}\,%
    \includegraphics[width=0.15\linewidth]{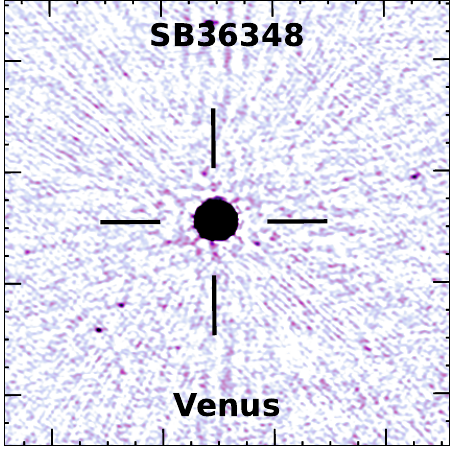}\,%
    \includegraphics[width=0.15\linewidth]{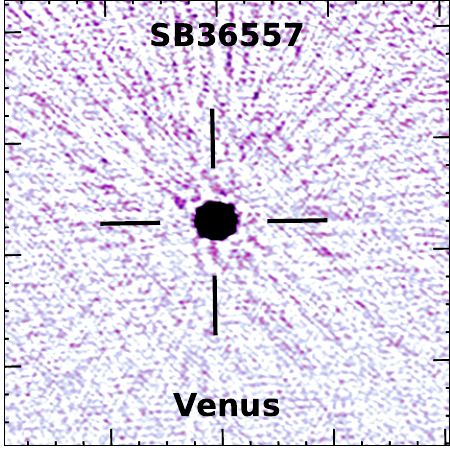}\\%
    \includegraphics[width=0.15\linewidth]{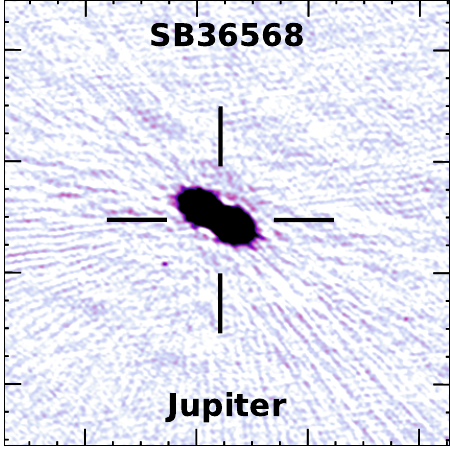}\,%
    \includegraphics[width=0.15\linewidth]{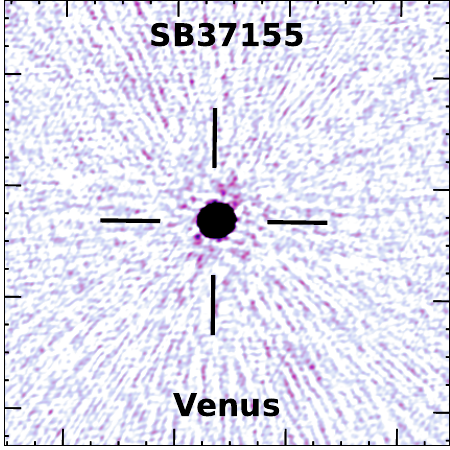}\,%
    \includegraphics[width=0.15\linewidth]{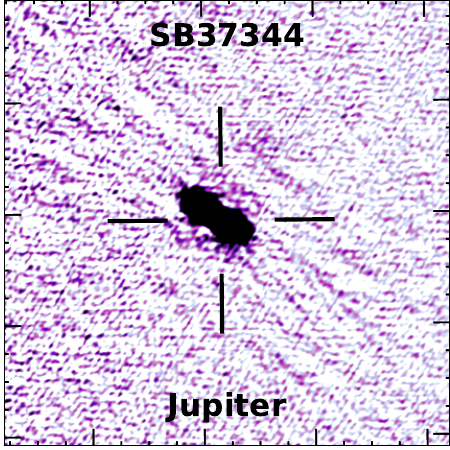}\,%
    \includegraphics[width=0.15\linewidth]{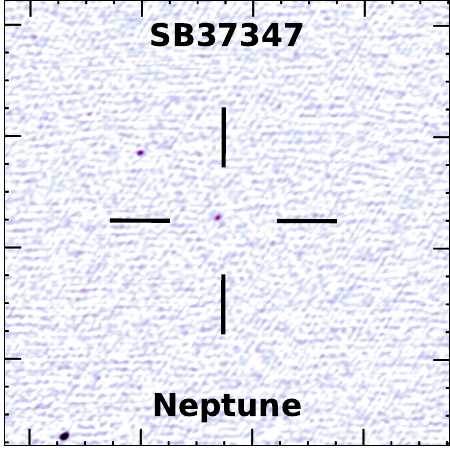}\,%
    \includegraphics[width=0.15\linewidth]{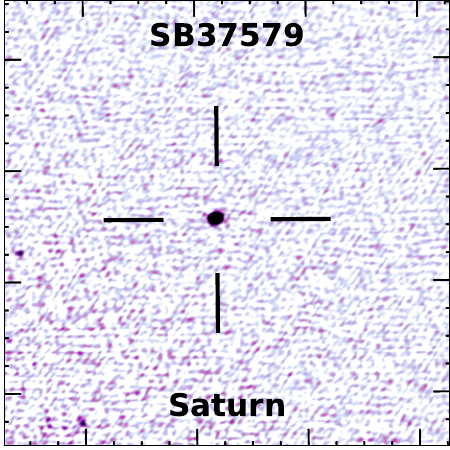}\,%
    \includegraphics[width=0.15\linewidth]{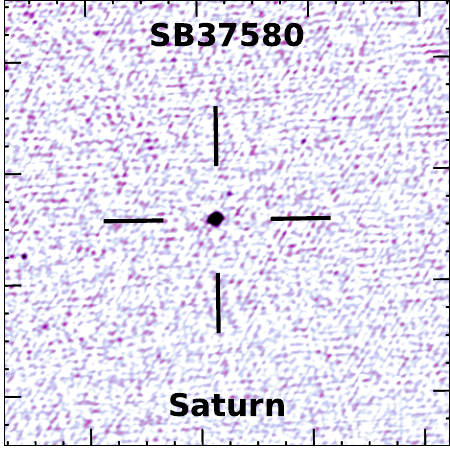}\\%
    \caption{\label{fig:planets} Solar system planets as they appear in the RACS-high images, in order of date observed (i.e. by SBID). We exclude SB37587, although it contains Uranus, because it is at the noisy edge of the image and not detected above $5\sigma_\text{rms}$.}
\end{figure*}

RACS observations show a time-dependent brightness scale fluctuation {when comparing source flux densities to external radio surveys} \citepalias[for both RACS-low and RACS-mid;][]{racs1,racs-mid}. For RACS-mid, we corrected this by using the SUMSS and NVSS cross-matched source lists, taking the \corrs{central 2 deg} of the image to avoid areas that might be affected by the---at the time---uncertain primary beam models. For RACS-mid observations that did not have cross-matches to SUMSS or NVSS, we used an elevation-dependent polynomial model derived from observations with SUMSS and/or NVSS cross-matches. RACS-high shows similar brightness scale fluctuations {with respect to the NVSS scaled to 1656\,MHz, with a median ratio of 1.05 (ranging from 0.89 to 1.27) across all observations above $\delta_\text{J2000} > -40\degr$.}

To highlight the periodicity of the fluctuations, we construct a Lomb--Scargle \citep[][see also \citealt{VanderPlas2018}]{Lomb1976,Scargle1982} periodogram of the average flux density ratio \corrs{(weighted by the signal-to-noise ratio, $\sigma$)} of sources cross-matched to the NVSS catalogue \corrs{(for each SBID, as described in Section~\ref{sec:imaging})}. Figure~\ref{fig:ls} shows the periodogram, which highlights a number of periodic features, with the most significant feature at $\approx 1$\,day. {We suspect the fluctuations in the brightness scale are related to temperature variation of the PAFs on each antenna. Such time-dependent fluctuations would normally be removed with more regular observations of a gain calibrator which is not done as part of ASKAP observing strategies.}

To model the time-dependent fluctuations, we use the 1-day sinusoidal model obtained from the Lomb--Scargle algorithm. We also see separate bandpass calibrator-dependent offsets which are not captured in the sinusoidal model. To account for this, the model is scaled by the mean ratio for each bandpass calibrator. For the three bandpass calibrators that only support SBIDs with no NVSS cross-matches (i.e. in the Southern Hemisphere) we apply the overall mean model value of 1.045. We note that we have assumed a spectral index \footnote{\corrs{We define the spectral index, $\alpha$, via $S_\nu \propto \nu^\alpha$, for a flux density $S_\nu$ at frequency $\nu$.}} of $\alpha=-0.8$ to scale the NVSS flux density measurements to \corrs{1\,655.5}\,MHz. Assuming other values between $-0.7$ and $-0.9$ would vary the overall mean value by $\approx 3$\,\%. {The difference in frequency between RACS-high and SUMSS (\corrs{1\,656} and 843\,MHz) makes it challenging to use a more southern widefield survey as an absolute flux density reference as the choice of spectral index, $\alpha$, becomes more significant in scaling the flux density measurements.}

The median flux density ratios for each SBID with respect to the NVSS are shown in the top panels of Figure~\ref{fig:fluxscale}. The 1-day sinudoidal model after scaling based on bandpass-specifc weighted-means is shown as a red line line on the top panel of Figure~\ref{fig:fluxscale}. The bottom panels of Figure~\ref{fig:fluxscale} show the ratios after applying the model factors to each SBID. 
The standard deviation, $\sigma$, is reduced from 0.061 to 0.034 after application of the model. As a point of comparison, only applying the bandpass-specific weighted means as a correction factor results in $\sigma=0.042$. Using the period model allows us to apply a reasonable correction factor to the SBIDs that do not have NVSS cross-matches. 

The correction factors are \textit{only} applied to images prior to the source-finding process described in the following sections. This means that the observation-specific images in CASDA are not scaled. The imaging data products produced after application of brightness scale correction factors will also be discussed in the following sections. 

\subsection{SBID selection}\label{sec:selection}

Some fields were observed multiple times due elevated root-mean-square (rms) noise or transiting Solar System bodies. Re-observations usually resulted in lower rms noise, but this sometimes came at the cost of coarser angular resolution due to a lower elevation pointing or loss of an outlying antenna. There are a total of 91 fields (of 1\,493) observed twice, and for construction of the catalogue, we defined a `best' observation for any duplicated field using a similar decision tree that was used for RACS-mid \citep[see Section 2.1 in][]{racs-mid2}. We relax the previous `peel' criterion (as all SBIDs undergo peeling that need it) and introduce instead a check for Solar System planets within the image. Solar System planets observed in RACS-high images are shown in Figure~\ref{fig:planets}. We also relaxed the minimum integration time as no RACS-high observations were less than 14.9\,min. The criteria are therefore \begin{enumerate}
    \item If both SBIDs feature planets within the images, we selected the SBID with the planet closest to the image edge,
    \item Else if one SBID features a planet within the image, we selected the SBID without a planet,
    \item Else if the difference in PSF major axes is $>2$\,arcsec between duplicate SBIDs, then we selected the SBID with the smaller PSF major axes,
    \item Else we selected the SBID with the lowest median rms noise.
\end{enumerate}

SBIDs not selected for the catalogue were still validated and released on CASDA and can be used for science. \corrs{Selected SBIDs are highlighted on Figure~\ref{fig:flag_summary}, which shows that SBIDs that were not selected often had a larger fraction of data flagged (hence the re-observation).}

\subsection{Full sensitivity images}\label{sec:mosaic}

Following RACS-low and RACS-mid, we use \texttt{SWarp} \footnote{\url{https://www.astromatic.net/software/swarp/}.} \citep{swarp} to construct a `full-sensitivity' mosaic of each field, combining the neighbouring fields to remove the areas of lowered sensitivity due to primary beam attenuation. As part of this process, the same weights used for mosaicking the individual PAF beam images are used again here. As with RACS-mid, we also convolve all images for each field mosaic to a common resolution using \texttt{beamcon\_2D} \footnote{\url{https://github.com/AlecThomson/RACS-tools}, which makes use of the \texttt{radio-beam} (\url{https://github.com/radio-astro-tools/radio-beam}) package to find the common PSF for all images.}. The field \texttt{RACS\_0526-73} required extra manual padding for both the PSF minor and major axes when finding the common PSF size, resulting in a final angular resolution of $19^{\prime\prime}\times 12^{\prime\prime}, 40.26^\circ$. 

\subsection{Source finding and creating the catalogue}\label{sec:sf}

For source-finding on the full sensitivity mosaic images, we again use \texttt{PyBDSF}\footnote{\url{https://github.com/lofar-astron/PyBDSF}.} \citep{pybdsf} following \citetalias{racs2}. Source-finding settings remain identical to those used for RACS-mid \citepalias{racs-mid2}: the source-finding threshold is $5\sigma_\text{rms}$, with $\sigma_\text{rms}$ calculated internally by \texttt{PyBDSF} with a grid and box size of $(15\,\theta_\text{min}, 3\,\theta_\text{min}$), scaling with the PSF minor axis ($\theta_\text{min}$) for each image. \texttt{PyBDSF} produces a list of 2-D Gaussian components and a `source' list that contains groups of connected components. For RACS-mid, we regrouped components for sources that had a potentially unnecessary number of components---i.e.\,a source that is close to a point source containing more than one component (section 2.3 in \citetalias{racs-mid2}). We repeat this process here, finding up to $\approx10\%$ of components are needing to be regrouped.

The individual source lists are then merged to construct the all-sky catalogue. As the full-sensitivity mosaics contain a significant overlap, when the source lists are merged we cross-match the lists and consider any source with a match in a different list within the mean measured source major axis to be a duplicate. For any duplicates, we only keep the source that lies closest to the image centre. The Gaussian component lists that are also produced by \texttt{PyBDSF} are concatenated, then matched using unique source identifiers in the source catalogue. The final source catalogue contains 2\,677\,509 sources, and the Gaussian component catalogue contains 3\,526\,674 components.

Appendix~\ref{app:columns} records the table columns for both the source (Table~\ref{tab:columns:source}) and Gaussian component (Table~\ref{tab:columns:component}) catalogues and are a subset of the columns in the RACS-mid catalogues. Three observation-specific columns present in the RACS-mid catalogues are not provided for RACS-high (SBID, scan length, and scan start) as all of the full-sensitivity images have contributions from multiple observations. For compactness, we also opt to remove the Galactic latitude and longitude columns as they can be derived from existing columns. The column for the brightness scale uncertainty, $\xi_\text{scale}$, is also removed as it has a single value over the whole survey (see Section~\ref{sec:validation:flux}).

The properties of the source catalogue are discussed in the following sections and a summary is provided in Table~\ref{tab:catalogue_summary}.

\section{An assessment of the catalogue and images}

\begin{table}[t]
\begin{threeparttable}
    \centering
    \caption{\label{tab:catalogue_summary} A summary of properties of the RACS-high catalogue.}
    \begin{tabular}{l l c}\toprule
        $N_\text{sources}$ & 2\,677\,509 & Section~\ref{sec:sf} \\
        $N_\text{components}$ & 3\,526\,674 & Section~\ref{sec:sf} \\
        Median $\theta_\text{major}$ & 11\farcs8 (7\farcs4--40\farcs2) & Section~\ref{sec:resolution} \\
        Median $\theta_\text{minor}$ & 8\farcs1 (6\farcs1--16\farcs5) & Section~\ref{sec:resolution} \\
        Median $\sigma_\text{rms}$ & $195_{-32}^{+43}$\,\textmu Jy\,PSF$^{-1}$ & Section~\ref{sec:tile_sensitivity} \\ 
        $\xi_\text{scale}$ & 10\,\% & Section~\ref{sec:validation:flux} \\ 
        $\Delta\delta_\text{J2000}$ \tnote{a} & $-0\farcs2 \pm 1\farcs 0$  & Section~\ref{sec:astrometry} \\
        $\Delta\delta_\text{corr,J2000}$ \tnote{b} & $0\farcs 0 \pm 0\farcs 7$ & Section~\ref{sec:astrometry} \\
       $ \Delta\alpha_\text{J2000}$ \tnote{a} & $0\farcs 0 \pm 0\farcs 6$ & Section~\ref{sec:astrometry} \\\bottomrule 
    \end{tabular}
\begin{tablenotes}[flushleft]
{\footnotesize \item[a] Derived from ICRF3 cross-matches. \item[b] After applying polynomial correction model.}
\end{tablenotes}
\end{threeparttable}
\end{table}

\begin{figure*}[p]
\centering
\includegraphics[width=1\linewidth]{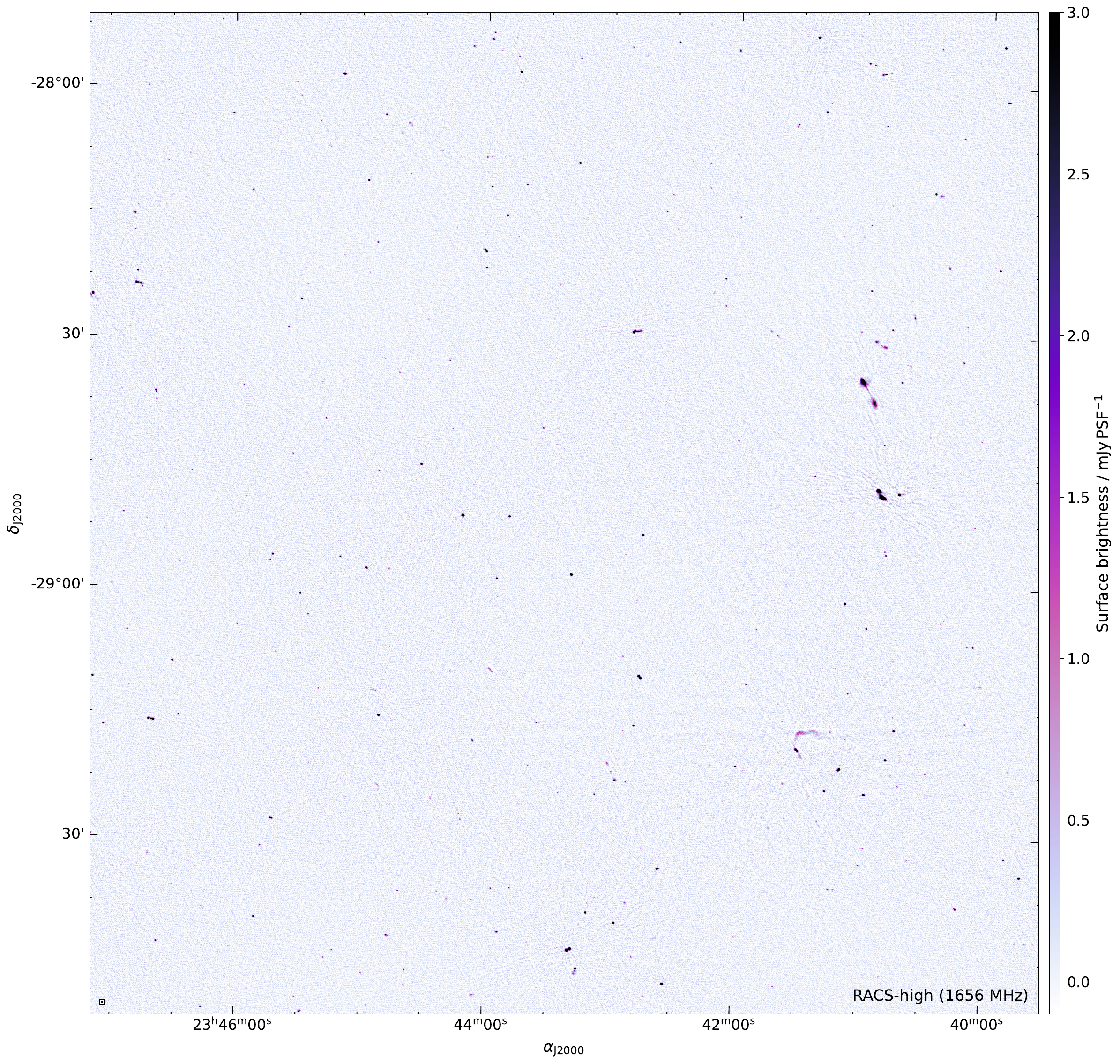}
\caption{\label{fig:largesky} An example $2\degr \times 2\degr$ cutout from a RACS-high full-sensitivity image. \corrs{The image has an angular resolution of $7\farcs4 \times 6\farcs2$, and a median rms noise of $\sigma_\text{rms} = 175$\,\textmu Jy\,PSF$^{-1}$.}}
\end{figure*}

\begin{figure*}[p]
    \centering
    \begin{subfigure}[b]{1\linewidth}
    \includegraphics[width=1\linewidth]{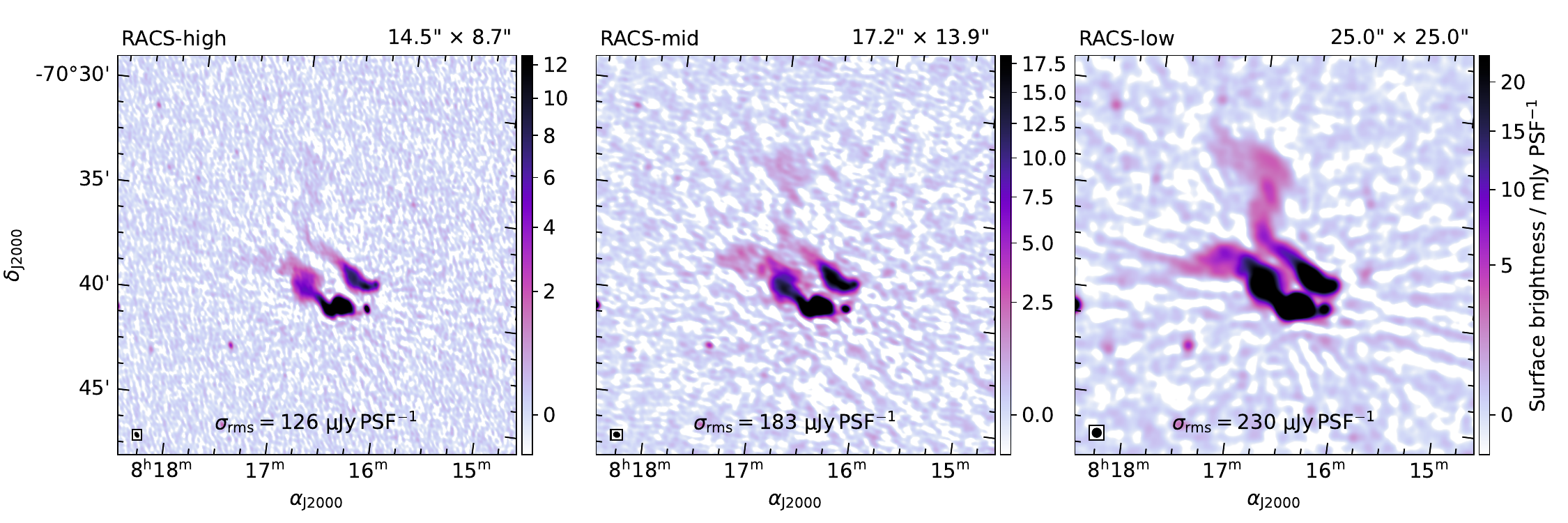}
    \caption{\label{fig:examples:pksj0816} PKS~J0816$-$7039.}
    \end{subfigure}\\%
    \begin{subfigure}[b]{1\linewidth}
    \includegraphics[width=1\linewidth]{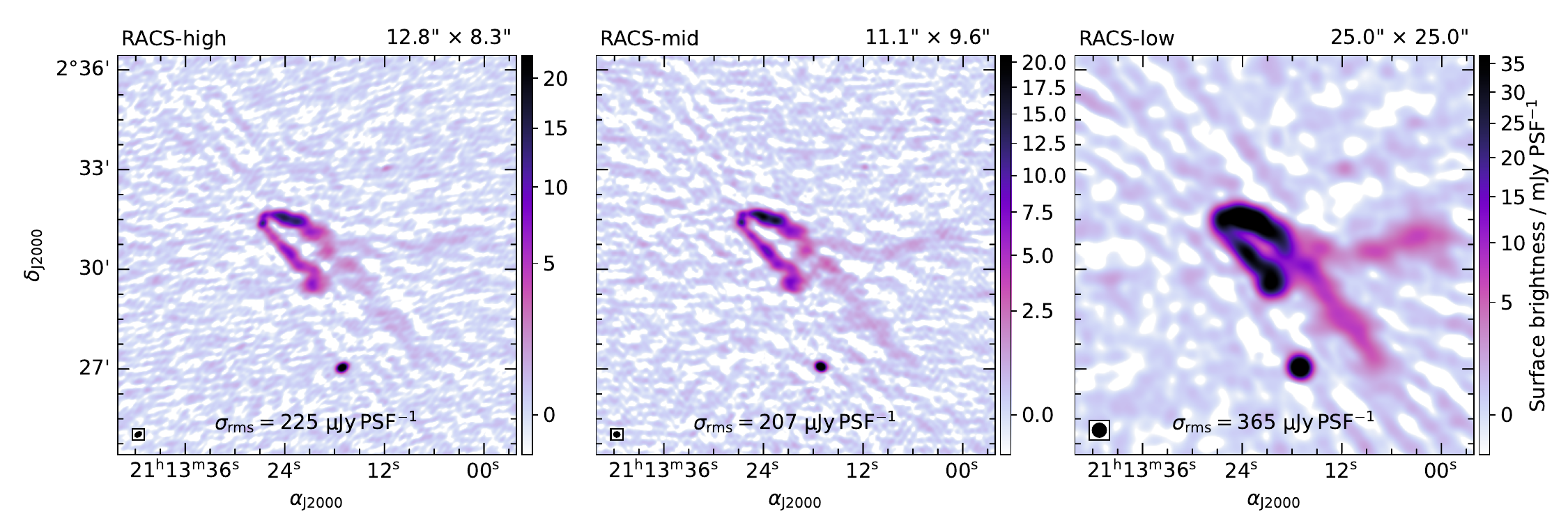}
    \caption{\label{fig:examples:pks2110} PKS~J2113$+$0230.}
    \end{subfigure}\\%
    \begin{subfigure}[b]{1\linewidth}
    \includegraphics[width=1\linewidth]{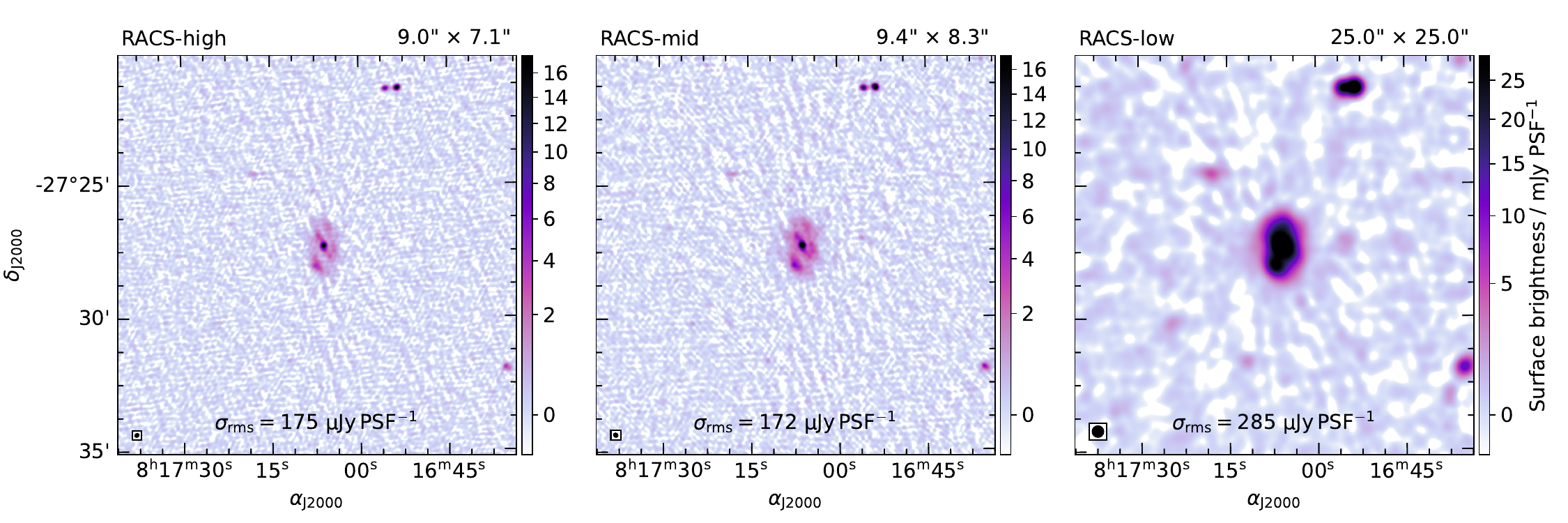}
    \caption{\label{fig:examples:ngc2559} NGC~2559.}
    \end{subfigure}\\%
    \caption{\label{fig:examples}Example cutouts from the RACS-high full-sensitivity images (\emph{left)} with comparison to the equivalent cutouts from RACS-mid (\emph{centre}) and RACS-low (\emph{right}). All colour scales used a square-root stretch in the range $[-1, 100]\sigma_\text{rms}$. \corrs{The shape of the PSF for each image is shown as an ellipse in the bottom left of each panel, and the major and minor axes are reported at the top of each panel.}}
\end{figure*}

\begin{figure}[t]
    \centering
    \includegraphics[width=1\linewidth]{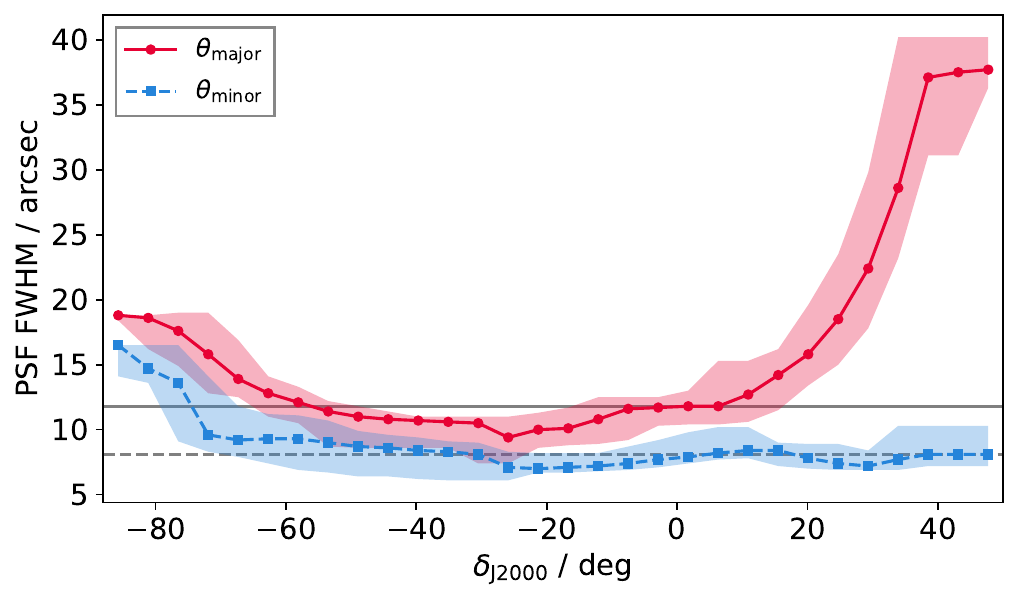}
    \caption{\label{fig:psf} The FWHM of the PSF major ($\theta_\text{major}$) and minor ($\theta_\text{minor}$) axes, binned as a function of declination. The Shaded regions show the range of values in each bin, and the horizontal, gray lines are drawn at the median values.}
\end{figure}

\begin{figure*}[t]
\centering
\includegraphics[width=1\linewidth]{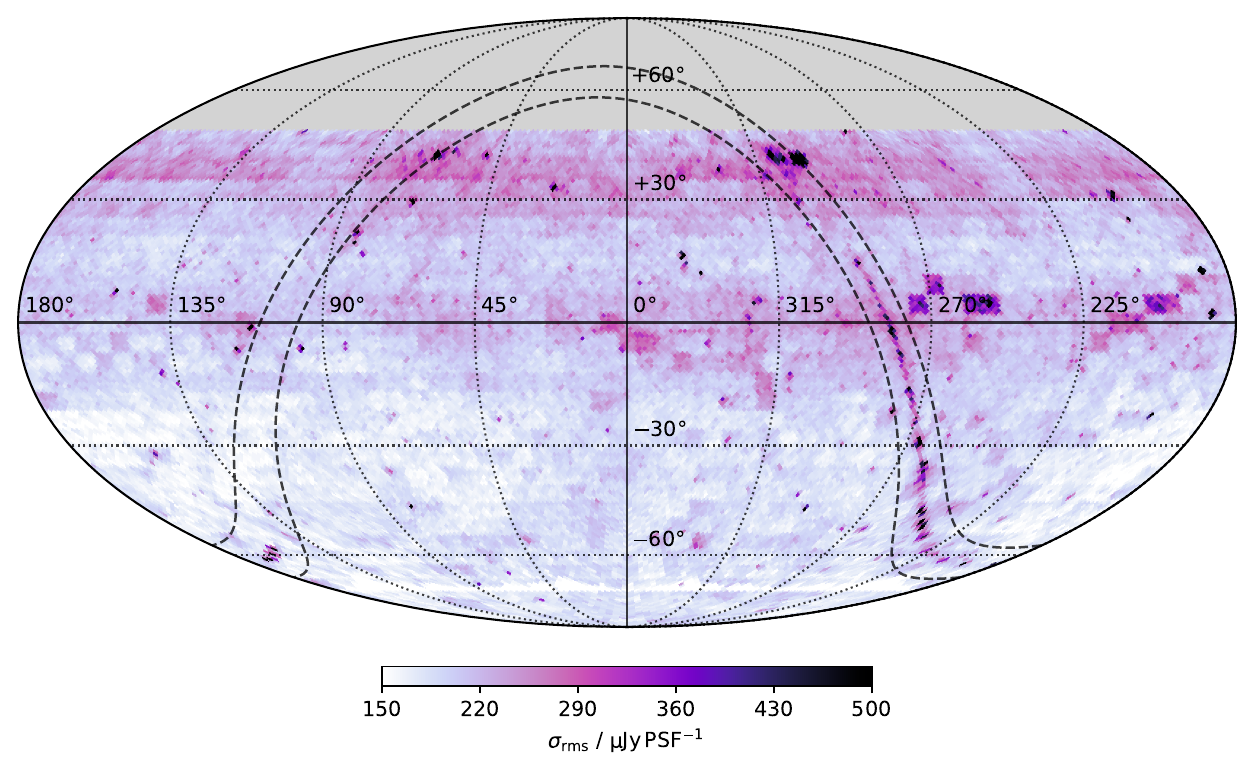}
\caption{\label{fig:rms} The HEALPix-binned rms noise as reported at source positions in the RACS-high catalogue.}
\end{figure*}

\begin{figure*}[t]
    \centering
    \begin{subfigure}[b]{0.5\linewidth}
    \includegraphics[width=1\linewidth]{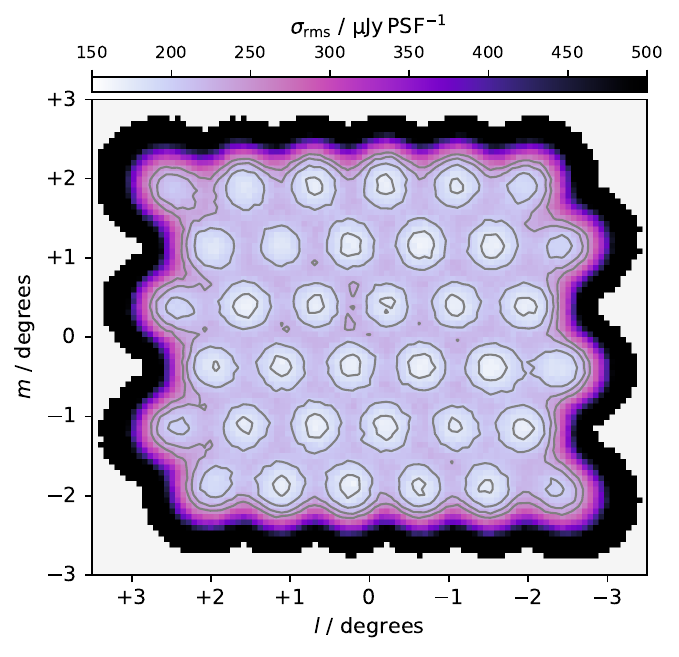}
    \caption{\label{fig:tile_sensitivity:sbid}}
    \end{subfigure}%
    \begin{subfigure}[b]{0.5\linewidth}
    \includegraphics[width=1\linewidth]{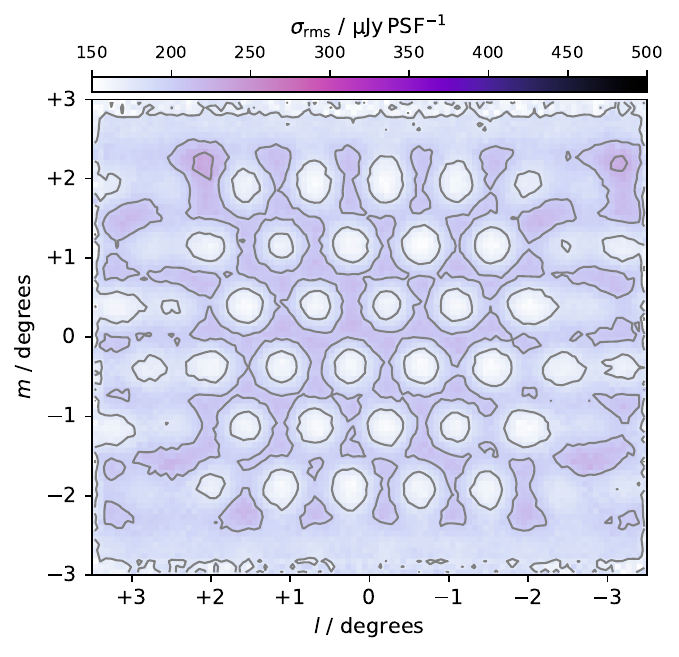}
    \caption{\label{fig:tile_sensitivity:full}}
    \end{subfigure}%
    \caption{\label{fig:tile_sensitivity} Measured median rms noise across the individual SBID tiles \subref{fig:tile_sensitivity:sbid} and the full-sensitivity mosaicked images \subref{fig:tile_sensitivity:full} as function of $(l,m)$ in the image reference frame. Median rms is computed in $\approx 3.6$-arcmin$^{2}$ cells. Contours are drawn at $[150, 175, 200, 225, 250]$\,\textmu Jy\,PSF$^{-1}$.}
\end{figure*}

\begin{figure}
    \centering
    \includegraphics[width=1\linewidth]{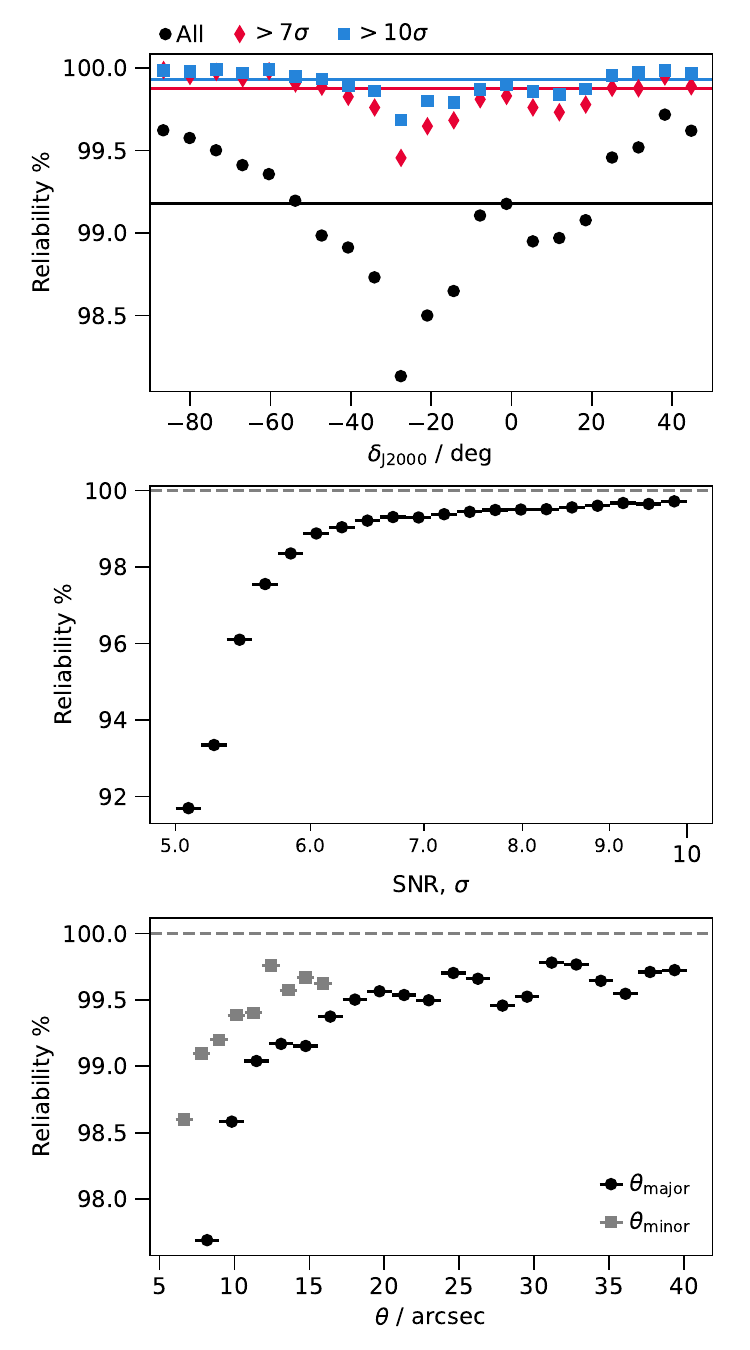}
    \caption{\label{fig:reliability} Reliability, binned as a function of declination (\emph{top}), $\sigma$ (\emph{middle}), and PSF major/minor axis size (\emph{bottom}). In the top panel, we show three $\sigma$ bins: no $\sigma$ cut (black circles), $>7\sigma$ (red diamonds), and $>10\sigma$ (blue squares), with horizontal lines drawn at median values. The shaded, \corrs{grey} horizontal lines in the middle and bottom panel are drawn at 100\%. In the middle and bottom panels the horizontal bars drawn with the markers show the bin width (these are removed from the top panel for clarity).}
\end{figure}

\begin{figure}[t]
    \centering
    \includegraphics[width=1\linewidth]{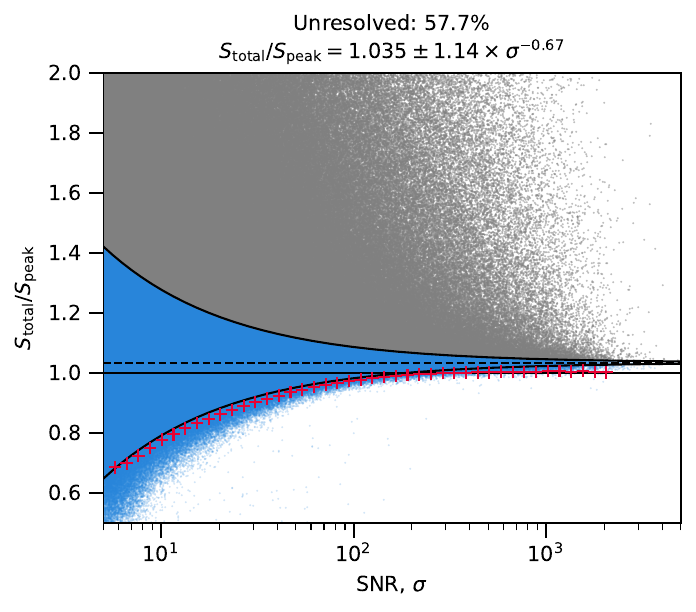}
    \caption{\label{fig:resolved} The ratio of total flux density ($S_\text{total}$) to peak flux density ($S_\text{peak}$) for sources as a function of $\sigma$. The dashed black line is the median ratio, and the solid, curved black lines indicate the fitted envelope that represents the  distribution of $S_\text{total}/S_\text{peak}$ for point sources. The red crosses indicate the binned 5$^\text{th}$ percentile, used to fit the envelope (Equation~\ref{eq:ip}). The sources are coloured by their \texttt{Flag} value, as described in the text, \corrs{unresolved sources (\texttt{0}) are blue, and resolved sources (\texttt{1}) are grey.}}
\end{figure}

\begin{figure}
    \centering
    \begin{subfigure}[b]{1\linewidth}
    \includegraphics[width=1\linewidth]{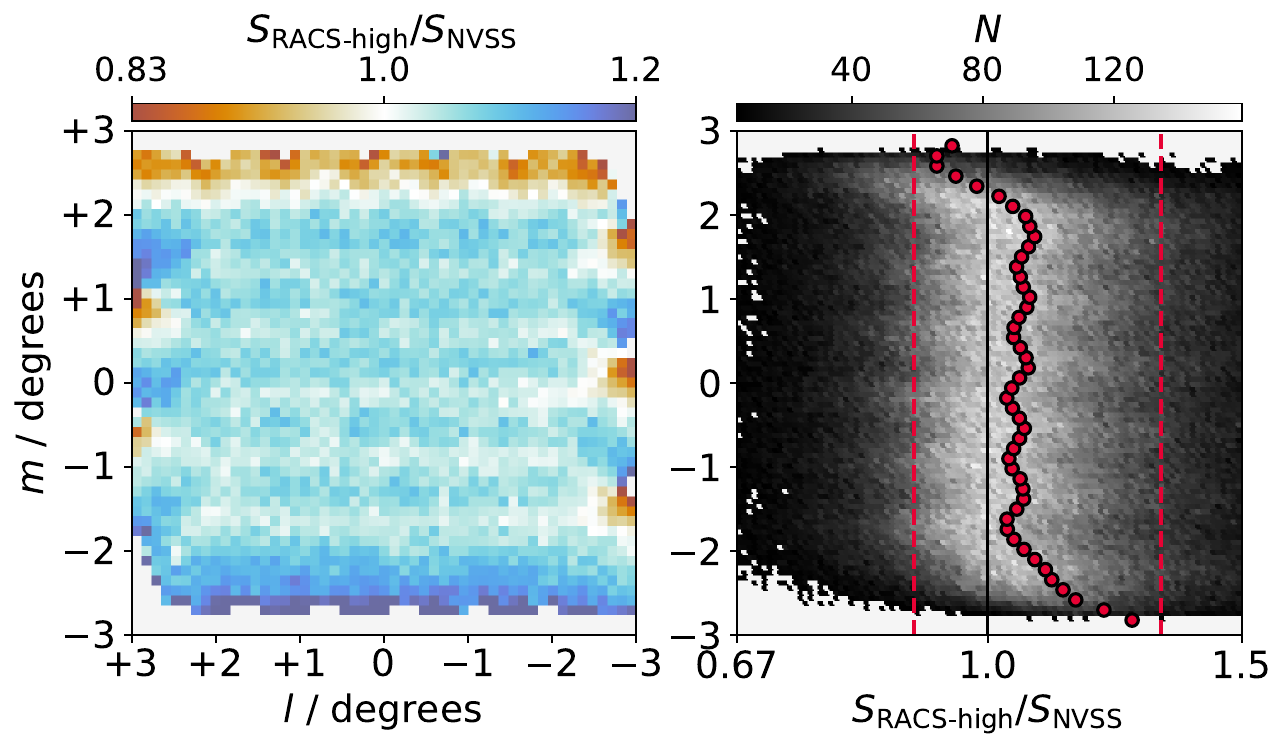}
    \caption{\label{fig:tileflux:pre1} Pre-mosaic images, no scaling.}
    \end{subfigure}\\%
    \begin{subfigure}[b]{1\linewidth}
    \includegraphics[width=1\linewidth]{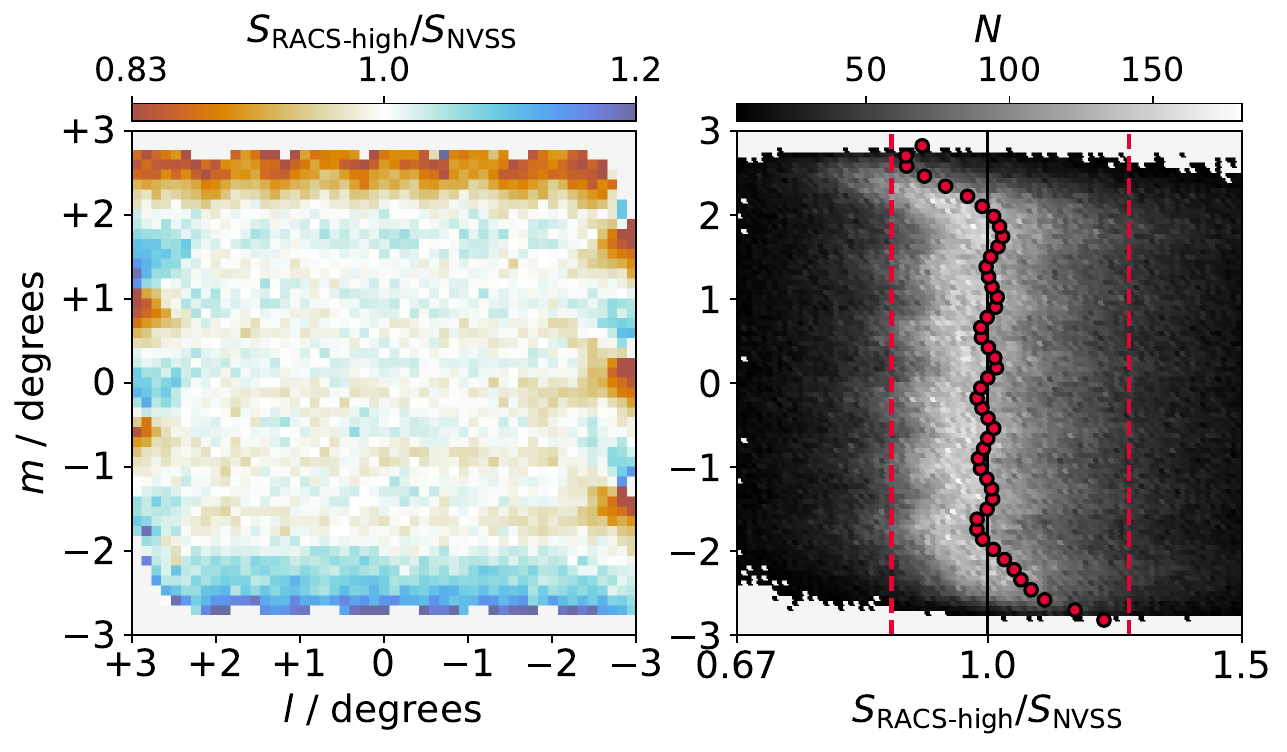}
    \caption{\label{fig:tileflux:pre2} Pre-mosaic images, after scaling.}
    \end{subfigure}\\%
    \begin{subfigure}[b]{1\linewidth}
    \includegraphics[width=1\linewidth]{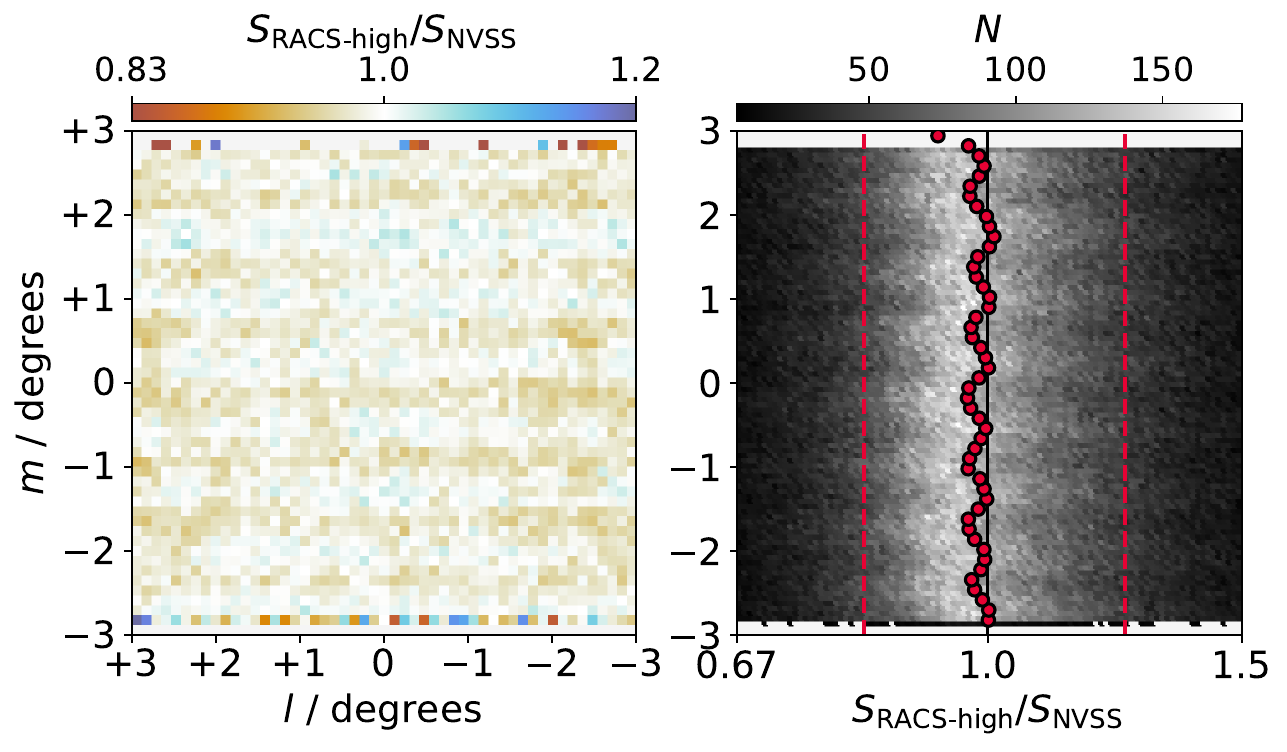}
    \caption{\label{fig:tileflux:post} Post-mosaic images.}
    \end{subfigure}\\%
    \caption{\label{fig:tileflux} The binned, median flux density ratio between RACS-high and the NVSS (used for absolute brightness scaling), as a function of position across the images. \ref{fig:tileflux:pre1} shows the brightness scale in individual images prior to scaling (what is available on CASDA). \ref{fig:tileflux:pre2} shows the individual images after applying the scaling factors derived from the NVSS (Section~\ref{sec:norm}) which are used for generating the full-sensitivity mosaic images, shown in \ref{fig:tileflux:post}. The right panels show the source density as a function of image coordinate $m$ and the flux density ratio, highlighting the ripples across the images. The red markers in the right panels show median ratios in bins as a function of $m$, with the vertical dashed red lines showing the 16-th and 84-th percentiles overall.}
\end{figure}

\begin{figure*}[t]
    \centering
    
    \begin{subfigure}[b]{0.5\linewidth}
    \includegraphics[width=1\linewidth]{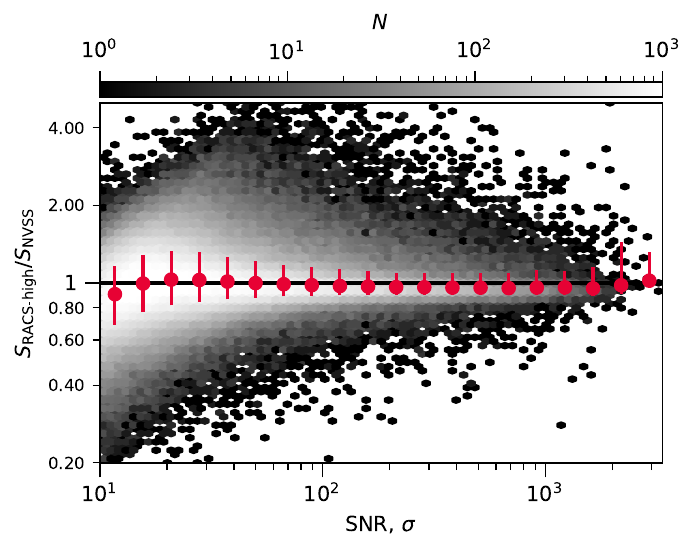}
    \caption{\label{fig:fluxsnr:nvss} RACS-high / NVSS.}
    \end{subfigure}%
    \begin{subfigure}[b]{0.5\linewidth}
    \includegraphics[width=1\linewidth]{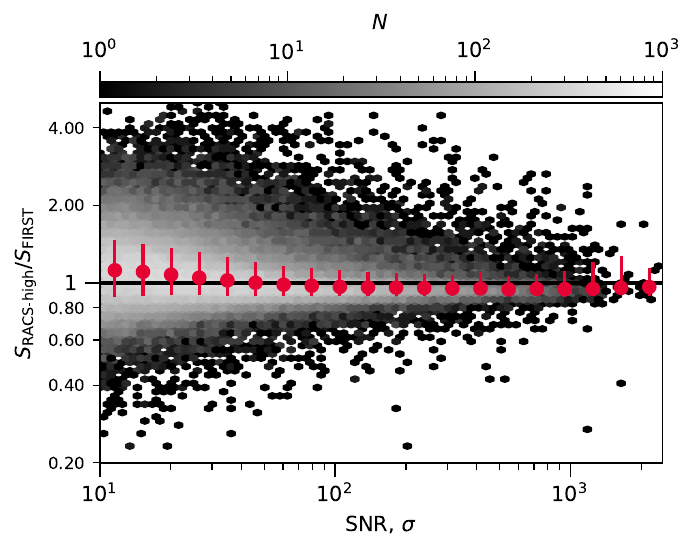}
    \caption{\label{fig:fluxsnr:first} RACS-high / FIRST.}
    \end{subfigure}%
    \caption{\label{fig:fluxsnr} Hexagonal binned flux density ratios between RACS-high and the NVSS \subref{fig:fluxsnr:nvss} and FIRST \subref{fig:fluxsnr:first}, after scaling to \corrs{1\,655.5}\,MHz, as a function of $\sigma$. The red circles show medians in bins as a function of \corrs{$\sigma$} with errors drawn from the 16-th and 84-th percentiles. The solid black line is drawn at a ratio of 1.}
\end{figure*}

\begin{figure}[t]
    \centering
    \includegraphics[width=1\linewidth]{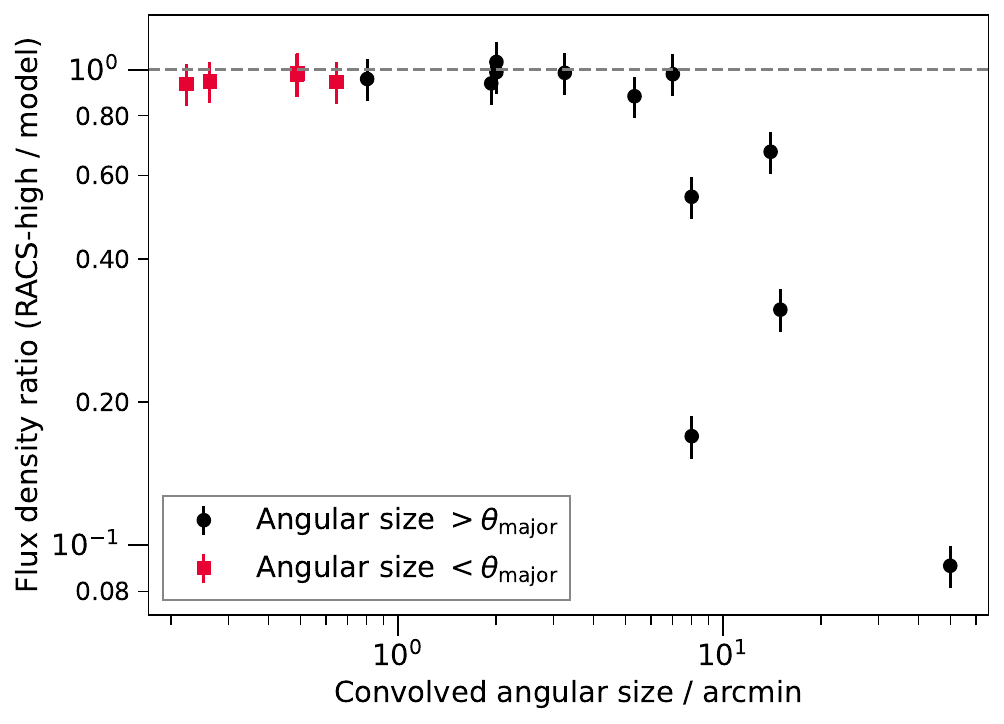}
    \caption{\label{fig:pbscale} Ratio of integrated flux density measured in the full-sensitivity RACS-high images to the model flux densities of calibrator sources from \citet{Perley2017} and \citet{rey94}, as a function of the convolved angular size of the source. The dashed, horizontal line indicates a ratio of 1. The two points with the lowest ratios are Taurus~A and Fornax~A, significantly resolved out of the RACS-high images. \corrs{Black circles indicate sources with an angular size greater than the PSF major axis ($\theta_\text{major}$) from the image within it is measured, and red squares indicate sources with an angular size less than $\theta_\text{major}$.}} 
\end{figure}

\begin{figure}[t]
    \centering
    \includegraphics[width=1\linewidth]{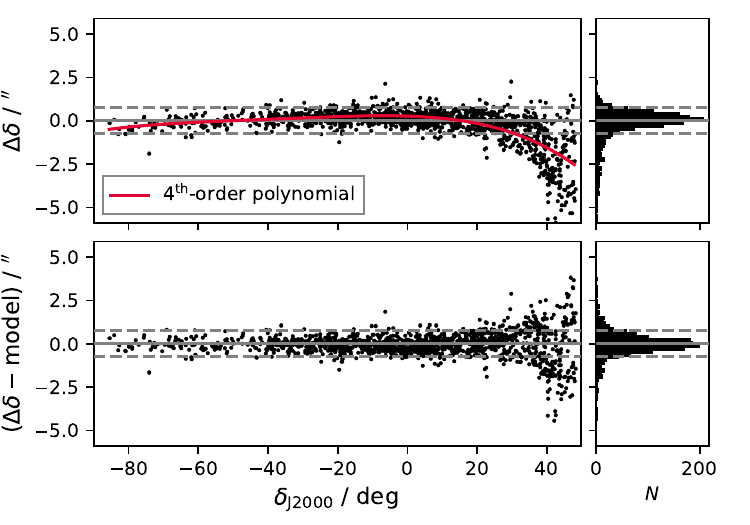}
    \caption{\label{fig:astrometry:dec} Declination offsets between RACS-high and the ICRF3 as a function of declination. \emph{Top panels.} The offsets prior to correction. The red, solid line shows the 4$^{th}$-th order polynomial fit to the offsets. \emph{Bottom panels.} The residuals after applying the polynomial model. The horizontal gray, solid lines are drawn at zero offset, and the horizontal grey, dashed lines are drawn at $\pm 0.75$~arcsec.}
\end{figure}

\begin{figure*}
    \centering
    \begin{subfigure}[b]{0.5\linewidth} 
    \includegraphics[width=1\linewidth]{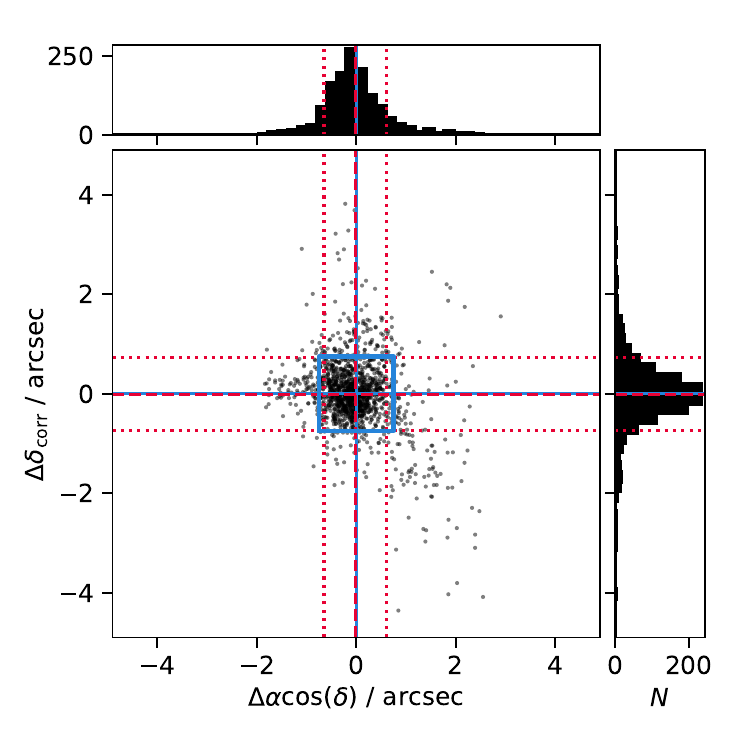}
    \caption{\label{fig:astrometry:icrf} RACS-high $-$ ICRF3.}
    \end{subfigure}%
    \begin{subfigure}[b]{0.5\linewidth} 
    \includegraphics[width=1\linewidth]{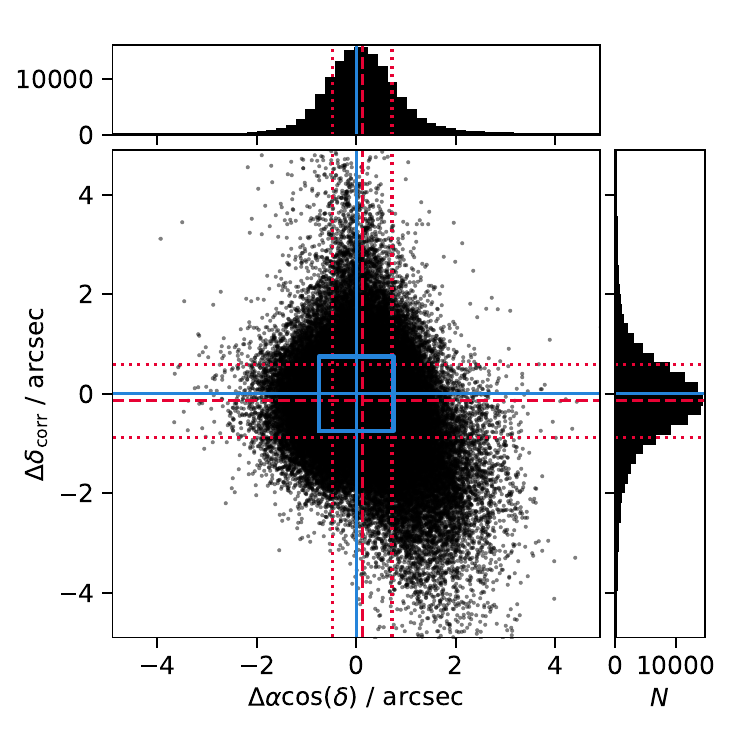}
    \caption{\label{fig:astrometry:first} RACS-high $-$ FIRST.}
    \end{subfigure}\\%
    \begin{subfigure}[b]{0.5\linewidth} 
    \includegraphics[width=1\linewidth]{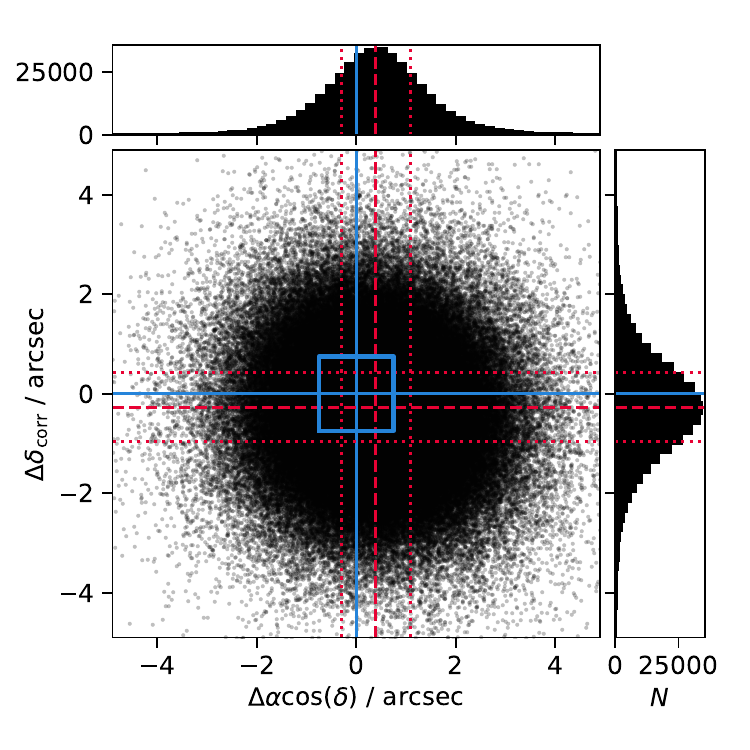}
    \caption{\label{fig:astrometry:low} RACS-high $-$ RACS-low.}
    \end{subfigure}%
    \begin{subfigure}[b]{0.5\linewidth} 
    \includegraphics[width=1\linewidth]{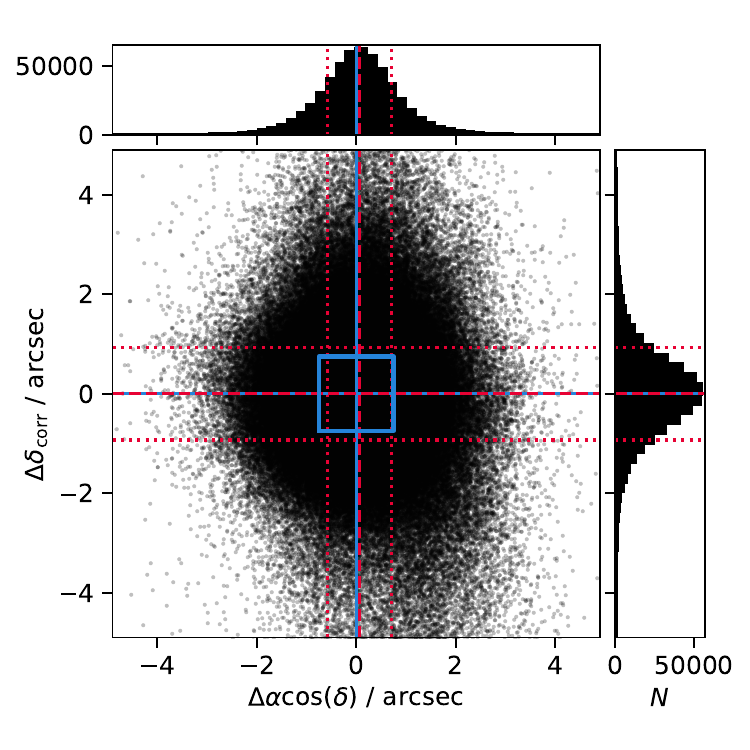}
    \caption{\label{fig:astrometry:mid} RACS-high $-$ RACS-mid.}
    \end{subfigure}\\%
    \caption{\label{fig:astrometry} Astrometric offsets between RACS-high (corrected declination measurement) and the ICRF3 \subref{fig:astrometry:icrf}, FIRST \subref{fig:astrometry:first}, RACS-low \subref{fig:astrometry:low}, and RACS-mid \subref{fig:astrometry:mid}. The solid blue lines are drawn at zero offset and the solid blue square shows the RACS-high pixel size (1.5\,arcsec). The red, dashed lines are drawn at the mean offset values, and the red, dotted lines are drawn at $\pm 1\sigma$ from the mean.}
\end{figure*}

\begin{table}[t]
    \centering
    \begin{threeparttable}
    \caption{\label{tab:astrometry} Mean and standard deviation of  $(\alpha_\text{J2000},\delta_\text{J2000})$ offsets for sources cross-matched between RACS-high and external catalogues, with offsets defined as $\Delta\delta = \delta_\text{RACS-high}-\delta_\text{survey}$ and $\Delta\alpha\cos\delta = (\alpha_\text{RACS-high} - \alpha_\text{survey})\cos\delta$.}
    \begin{tabular}{cc c c c}\toprule
         Survey & $N_\text{sources}$ & $\Delta\alpha\cos\delta$ & $\Delta\delta$ & $\Delta\delta_\text{corr}$ \\
          & & (arcsec) & (arcsec) & (arcsec) \\[0.5em]\midrule
         ICRF3 & 1\,415 & $-0.02 \pm 0.63$ & $-0.24 \pm 1.04$ & $-0.01 \pm 0.74$ \\
         FIRST & 125\,887 & $+0.12 \pm 0.71$ & $-0.47 \pm 1.22$ & $-0.15 \pm 0.92$ \\
         RACS-low & 394\,345 & $+0.39 \pm 1.06$ & $-0.24 \pm 1.05$ & $-0.30 \pm 1.06$ \\
         RACS-mid \tnote{a} & 537\,929 & $+0.06 \pm 0.81$ & $-0.14 \pm 1.29$ & $-0.00 \pm 1.16$ \\[0.5em]\bottomrule
    \end{tabular}
\begin{tablenotes}[flushleft]
    {\footnotesize \item[a] Cross-matched to the equivalent corrected declination measurement for RACS-mid.}
\end{tablenotes}

\end{threeparttable}
\end{table}

We encourage users to make use of the all-sky catalogue and the full-sensitivity mosaic images instead of the images and source lists available separately for each observation. The exception might be if there is a need for an epoch-specific source detection, a need for an image with the original brightness scale, or the highest possible resolution (see Section~\ref{sec:resolution}). The following assessment of data quality primarily focuses on these higher-order data products. A summary of the catalogue properties that are discussed on the following sections is provided in Table~\ref{tab:catalogue_summary}.

Figure~\ref{fig:largesky} shows a $2\degr \times 2\degr$ region from an example full-sensitivity image, showing a range of point sources and extended radio galaxies. In figure~\ref{fig:examples} we also show a selection of cutouts of sources from the RACS-high full-sensitivity images alongside the equivalent full-sensitivity cutouts from RACS-mid \citepalias{racs-mid2} and RACS-low \citepalias{racs2}. In the case of RACS-low, these images are convolved to a fixed $25\,\text{arcsec} \times 25\,\text{arcsec}$ angular resolution. The examples show two radio galaxies [PKS~J0816$-$7039, Figure~\ref{fig:examples:pksj0816}, and PKS~J2113$+$0230, \ref{fig:examples:pks2110}] and a star-forming spiral galaxy [NGC~2559, Figure~\ref{fig:examples:ngc2559}], highlighting a range of angular scales from compact radio cores, to larger-scale radio lobes.

\subsection{Angular resolution}\label{sec:resolution}

For the original per-SBID images, the median angular resolution is $8\farcs 6 \times 6\farcs 3$. The FWHM of the PSF major axis ($\theta_\text{major}$) is in the range $6\farcs 5$--$37\farcs 1$, and the minor axis ($\theta_\text{minor}$) is in the range $5\farcs 5$--$13\farcs 7$. These per-SBID images available through CASDA represent the highest-resolution RACS images available. The median resolution increases by $0\farcs1$ in both axes when considering SBIDs selected for mosaicking and catalogue work as many of the reobservations had a 1--2\arcsec coarser resolution.  

As described in Section~\ref{sec:mosaic}, the catalogue is constructed from the full-source-lists derived from the full-sensitivity images. Convolving images to a common resolution prior to mosaicking decreases the angular resolution on average and each full-sensitivity image has a different angular resolution. The resultant catalogue therefore  has a position-dependent angular resolution. Figure~\ref{fig:psf} shows the variation of the PSF, binned as a function of declination. The median PSF axes in the full-sensitivity images as reported in the catalogue at source positions is $\theta_\text{major} = 11\farcs8$ (ranging from $7\farcs4$--$40\farcs2$) and $\theta_\text{minor} = 8\farcs1$ (ranging from $6\farcs1$--$16\farcs5$). The angular resolution for the full-sensitivity images catalogue is therefore $\approx 3\arcsec \times 2\arcsec$ coarser than the per-SBID images, but still largely changes as a function of declination. Figure~\ref{fig:examples:pks2110} shows an example RACS-high image that has a similar angular resolution ($12\farcs8 \times 8\farcs3, 122.2\degr$), to the RACS-mid counterpart ($11\farcs1 \times 9\farcs6, 83.7\degr$). In that case, the original RACS-high image PSF is $8\farcs0 \times 6\farcs7, 74.0\degr$ and highlights an example where the original images may be more useful when requiring the highest-possible angular resolution. Figures~\ref{fig:examples:pksj0816} and \ref{fig:examples:ngc2559} show examples where the angular resolution is higher than the other RACS images as expected.

\subsection{Root-mean-square noise and sensitivity}\label{sec:tile_sensitivity}

Figure~\ref{fig:rms} shows the Hierarchical Equal Area isoLatitude Pixelation \citep[HEALPix;][]{Gorski2005} binned ($N_\text{side} = 64$) rms noise reported in the RACS-high catalogue (i.e. at source positions). The general noise properties are similar to RACS-mid, with noticeable increases in rms noise through parts of the Galactic Plane, around bright sources, and towards higher declination. There are also increases in rms noise around the celestial equator and certain fields, partially due to differences in RFI and subsequent flagging. The overall median rms noise as reported in the catalogue is $195_{-32}^{+43}$\,\textmu Jy\,PSF$^{-1}$.

RACS-high has the same observation setup as RACS-mid, except for the change in frequency. This includes the same arrangement and spacing of the 36 PAF beams. At the higher frequency, the primary beam shape has a smaller full-width at half maximum (FWHM) and the overlap between PAF beams is less than in the mid and low bands. The consequence of this is a comparative reduction of median sensitivity across the RACS-high images, and a noticeable sensitivity decrement in the spaces between PAF beams. 

In Figure~\ref{fig:tile_sensitivity} we show the median rms noise across the original images [\ref{fig:tile_sensitivity:sbid}] and the across the full-sensitivity mosaic images [\ref{fig:tile_sensitivity:full}]. The position-dependent rms noise is obtained from the individual source lists for each image (and are therefore measurements at source positions) and we obtain median values within $\approx 3.6$-arcmin$^2$ bins in $(l,m)$ coordinates defined in the reference frame of the individual images. Figure~\ref{fig:tile_sensitivity:sbid} shows the binned rms noise prior to mosaicking, highlighting the increase in noise towards the edge of the PAF footprint due to primary beam sensitivity roll-off. Figure~\ref{fig:tile_sensitivity:full} shows the binned rms noise after mosaicking, with the image edges now consistent with the interior of the images.

\subsection{Reliability}\label{sec:reliability}

An assessment of the general reliability of the catalogue and images is performed by multiplying the full-sensitivity images by $-1$ and re-running the source-finding using the same thresholds. This process mirrors that done for RACS-low (section 6.4 in \citetalias{racs2} and RACS-mid (section 4.5 in \citetalias{racs-mid2}) and for many other radio surveys \citep[e.g.][]{ijmf16,HurleyWalker2022}. This procedure assumes that noise is symmetric and that the number of sources found represents the number of artefacts likely found in the positive (normal) images. We construct a catalogue of negative sources, using the same source-list merger process we use for the main catalogue. We define the reliability as $(1 - N_\text{negative} / N_\text{positive})\times 100$\%, indicating the percentage of sources we can expect to be real. For this process, we exclude sources near the Galactic Plane ($|b|<5\degr$) and the following analysis is only relevant for non-Galactic regions of sky. 

Figure~\ref{fig:reliability} shows the reliability, binned as a function of declination ({top panel}), $\sigma$ ({middle panel}), and PSF major and minor axis size ({bottom panel}). The overall reliability is 99.18\,\%. We find the reliability decreases with higher angular resolution, from 97.69\,\% to 99.72\,\% in the highest- and lowest-resolution bins, respectively. \corrs{We find reliability decreases near declination $\approx -26\degr$, which corresponds to the area of sky with the highest angular resolution, and in the bottom panel of Figure~\ref{fig:reliability} we see reliability decreasing as angular resolution improves. A similar feature was observed for RACS-mid \citepalias[][]{racs-mid2}, and visual inspection of the images with high angular resolution shows clear residual artefacts around sources. These artefacts are above the source-finding thresholds used by \texttt{PyBDSF}. The rms calculations are already modified from default values within \texttt{PyBDSF} to depend on the PSF minor axis to account for these additional artefacts near bright sources (see Section~\ref{sec:sf}), though this does not reduce their detection completely. Reliability otherwise improves as a function of $\sigma$ as expected.} 

\subsection{Unresolved sources}\label{sec:resolved}

For RACS-mid a set of flags were applied to sources in the catalogues to note if a source is resolved, unresolved, or likely an artefact \citepalias[][see section 4.4]{racs-mid2}. We repeat this process for the RACS-high catalogue, excluding sources with Galactic latitudes $|b|<5\degr$, but find that the likely artefact population had minimal real artefacts so that flag is dropped from this catalogue. The flags are derived from the distribution of the ratio of source total flux density ($S_\text{total}$) to peak flux density ($S_\text{peak}$) as a function of source signal-to-noise ratio ($\sigma$). \corrs{As sources with $S_\text{total}/S_\text{peak}>1$ are expected to be resolved within some statistical variation, we define a function that captures the envelope encompassing the expected variation of $S_\text{total}/S_\text{peak}$ as a function of $\sigma$---e.g.\ equation 1 from \citetalias{racs2} (see section 5.2.1): \begin{equation}\label{eq:ipmodel}
    S_\text{total} / S_\text{peak} = \text{med}\left(S_\text{total} / S_\text{peak}\right) \pm A \sigma^B \, ,
\end{equation} 
We assume sources with $S_\text{total}/S_\text{peak} < 1$ are unresolved, and calculate the 100$-$95-th percentile in $\sigma$ bins for sources below the median ratio. We fit Equation~\ref{eq:ipmodel} to the binned values, and assuming the function is symmetric about the median ratio we find} \begin{equation}\label{eq:ip}
S_\text{total} / S_\text{peak} = 1.035 \pm 1.14 \times \sigma^{-0.67} \, .
\end{equation}
Sources with $S_\text{total}/S_\text{peak}$ above Equation~\ref{eq:ip} are considered resolved (with \texttt{Flag = 1} in the catalogue) and the remaining sources are considered unresolved (\texttt{Flag = 0}). Figure~\ref{fig:resolved} shows the $S_\text{total}/S_\text{peak}$ distribution as a function of $\sigma$ with the fitted envelope shown. We find 57.7\% of sources are considered unresolved (blue points in Figure~\ref{fig:resolved}) based on this definition. This population is used for comparisons to other surveys in the following sections.

\subsection{Brightness scale}\label{sec:validation:flux}

All image source lists are cross-matched to the NVSS, considering unresolved, isolated (no neighbours within 90\,arcsec), and $\sigma>10$ sources only. While we do not provide the brightness-scaled images prior to making the full-sensitivity mosaics, we show the effect of applying the brightness scale to the images in Figure~\ref{fig:tileflux}. Figure~\ref{fig:tileflux} shows the ratio of RACS-high to NVSS flux density (assuming $\alpha = -0.8$) as a function of their position in the image reference frame for images prior to brightness scaling [Figure~\ref{fig:tileflux:pre1}, Section~\ref{sec:norm}], after brightness scaling [Figure~\ref{fig:tileflux:pre2}], and after making full-sensitivity mosaics with the brightness-scaled images [Figure~\ref{fig:tileflux:post}]. By construction, the brightness scale normalisation shifts the overall median flux density ratio to $1.00_{-0.14}^{+0.25}$, from $1.05_{-0.16}^{+0.27}$ prior to scaling, which does not change appreciably after making the full-sensitivity mosaics. {A lingering brightness scale ripple is seen in all panels of Figure~\ref{fig:tileflux}, particularly noticeable towards the images edges in Figure~\ref{fig:tileflux:pre1} and \ref{fig:tileflux:pre2}. This under- and over-correction is believed to be caused by an as yet unresolved telescope pointing error.} These features result in $\approx 20$\,\% discrepancy between the brightness scales at the very edges of the original images, which is reduced after making the full-sensitivity mosaics. This can be seen in RACS-mid as well, with figure~7 from \citetalias{racs-mid} highlighting this for a single example PAF beam and primary beam model from RACS-mid. 

The flux density ratio varies marginally as a function of $\sigma$, shown in Figure~\ref{fig:fluxsnr} for both the NVSS and FIRST cross-matches. For $>100\sigma$ sources, the median ratio between RACS-high and the NVSS [Figure~\ref{fig:fluxsnr:nvss}] reduces to $0.96_{-0.06}^{+0.09}$. We see the same ratio for 100$\sigma_\text{rms}$ sources cross-matched to the Faint Images of the Radio Sky at Twenty Centimeters \citep[FIRST;][]{Becker95,White1997,hwb15} catalogue [Figure~\ref{fig:fluxsnr:first}]. Based on the quadrature sum of $4\%$ offset and the 84-th percentile range for the 100$\sigma$ sample, and the $2\%$ brightness scale uncertainty inherited from the NVSS \citep{ccg+98}, we suggest an overall brightness scale uncertainty $\xi_\text{scale} = 10$\,\% for the RACS-high catalogue and full-sensitivity mosaic images.

As a separate inspection of absolute brightness scale in the images, we compare integrated flux density measurements of a selection of bright calibrator sources from \citet{Perley2017}. We evaluate their models at 1655.5\,MHz and measure their integrated flux density above $2\sigma_\text{rms}$ within apertures that are 1.5 times the size reported by \citet{Perley2017} convolved with $\theta_\text{major}$. We also include a comparison with PKS~B1934$-$638, our bandpass and initial absolute flux calibrator, with a model from \citet{rey94}. Figure~\ref{fig:pbscale} shows the ratio of RACS-high integrated flux density  to the flux density extrapolated from the reported models, as a function of convolved angular size. As with RACS-mid, we see a clear decrease in flux density as a function of source size due to incomplete $(u,v)$ coverage. This effect is stronger for RACS-high: at higher frequency the shortest baselines correspond to larger angular scales, and with the enforced 100\,m baseline cut for all data during imaging we lose even greater sensitivity to extended sources. Despite this, we have general agreement with the model flux densities up to $\sim 3$\,arcmin for these bright sources.

\subsection{Astrometry}\label{sec:astrometry}

For assessment of the astrometric accuracy, we cross-match the RACS-high catalogue to the ICRF3, FIRST, RACS-low, and the primary RACS-mid catalogue. As with the flux density comparisons, to avoid positional errors due to extended radio features or confusion with nearby sources, we take only isolated and unresolved sources (using \texttt{Flag = 0}). Offsets in $(\alpha_\text{J2000},\delta_\text{J2000}$) are then defined as the RACS-high position minus the external catalogue position.  

As with RACS-mid, we find that the offsets in declination increase as a function of declination, and we provide a `corrected' declination measurement by fitting a declination-dependent polynomial model to the declination offsets. Following the process used for the RACS-mid catalogues, we try a range of polynomials up to 5-th order and use the Akaike Information Criterion \citep[AIC;][]{Akaike1974} to select an appropriate model. For RACS-mid, a 2-nd order polynomial model was used, but for RACS-high we find that a 4-th order polynomial model is selected based on the AIC, though shows an overall similar shape. The model is shown on Figure~\ref{fig:astrometry:dec}, and the declination offsets after applying the model are shown in the bottom panel. The polynomial model is 
\begin{multline}
    \Delta\delta = +(0.27 \pm 0.05) - (6.3 \pm 2.6)\times 10^{-3} \delta \\ - (5.4 \pm 0.5) \times 10^{-4} \delta^2 - (8.8 \pm 1.5) \times 10^{-6} \delta^3 \\ -(5.4\pm 1.8)\times 10^{-8} \delta^4 \, \text{arcsec} ,
\end{multline}
with $\delta = \delta_\text{J2000}$ in degrees, and uncertainties derived from random sampling assuming the mean and standard deviation offsets from the sample. The largest model offset in declination is $-2.5$\,arcsec. In the catalogue, the `corrected' declination measurements are provided as a separate column named \texttt{Dec\_corr} (with associated uncertainty \texttt{E\_Dec\_corr}). The standard declination measurement column (\texttt{Dec}) is not altered to remain consistent with source positions in the images from which sources are found.

In Figure~\ref{fig:astrometry} we show the $(\alpha,\delta_\text{corr})$ offsets for RACS-high and the ICRF3 \subref{fig:astrometry:icrf}, FIRST \subref{fig:astrometry:first}, RACS-low \subref{fig:astrometry:low}, and RACS-mid \subref{fig:astrometry:mid}. In all cases except for RACS-low, the scatter in offsets in declination are still more significant than in R.~A. even after correction, though the mean offset is closer to zero. In Table~\ref{tab:astrometry} we summarise the mean and standard deviation of the offsets between RACS-high and the external catalogues. The offsets in declination are also reported with and without the correction. We also report the offsets for FIRST as another predominantly Northern Sky survey. In all cases except for RACS-low the declination correction reduces the standard deviation and moves the mean closer to zero. RACS-low is not affected by this change as RACS-low has the same error within the data though due to way in which RACS-low was observed this effect presents in both $\alpha_\text{J2000}$ and $\delta_\text{J2000}$, related to the elevation of the observations. We consider the astrometric accuracy to be $0\farcs 6$\,arcsec in $\alpha_\text{J2000}$ and and $0\farcs 7$ in $\delta_\text{J2000}$ (or 1\farcs 0 without correction), based on 1-$\sigma$ from the distributions of the ICRF3 offsets.

While the declination-dependent correction does reduce the standard deviation of the offsets somewhat, the bulk of the astrometric uncertainty inherent to many ASKAP observations remains. We note that the pixel size for the RACS-high images is $1.5\,\text{arcsec} \times 1.5\,\text{arcsec}$, though as discussed in previous papers in this series (section~3.4.3 in \citetalias{racs1}, section~3.6 in \citetalias{racs-mid}, and section~4.10 in \citetalias{racs-mid2}), the lack of phase referencing during observations limits the astrometric accuracy such that increasing $\sigma$ does not improve astrometric accuracy as much as expected.

\section{An initial look at RACS spectral modelling}\label{sec:spectra}

\begin{figure}[t]
    \centering
    \includegraphics[width=1\linewidth]{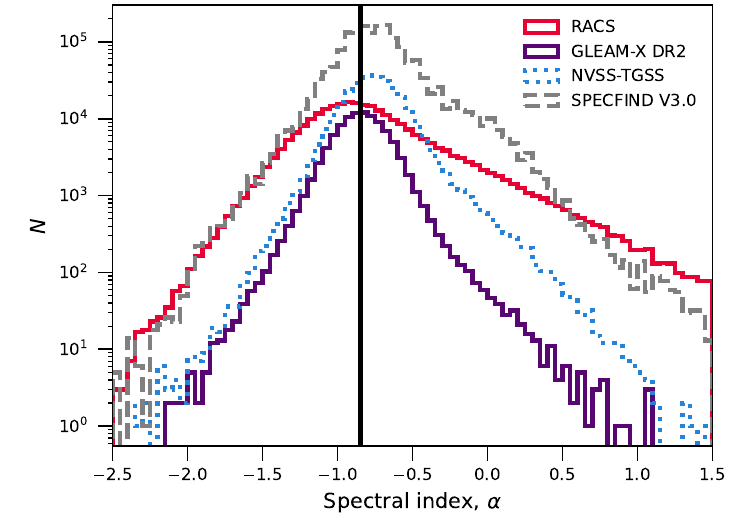}
    \caption{\label{fig:sed:hist} Histogram of spectral index, $\alpha$, from the cross-matched RACS catalogue, the GLEAM-X DR2 catalogue \citep{Ross2024}, the NVSS-TGSS spectral index catalogue \citep{deGasperin2018}, and SPECFIND~V3. \citep{Stein2021}. The solid black line shows the median $\alpha$ for RACS.}
\end{figure}

\begin{figure*}[t]
    \centering
    \begin{subfigure}[b]{0.33\linewidth}
    \includegraphics[width=1\linewidth]{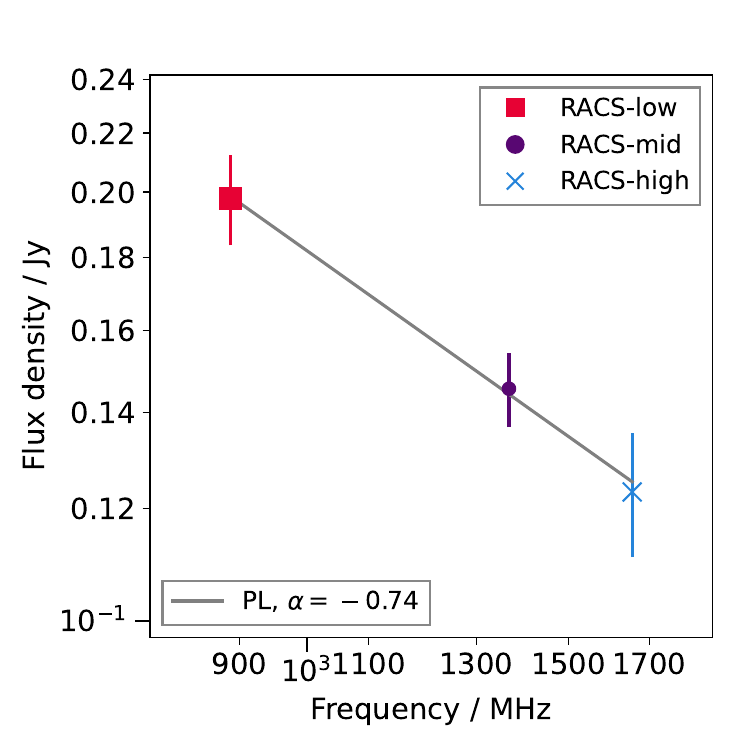}
    \caption{\label{fig:seds:j000} RACS-HIGH1~J000059$-$66322.}
    \end{subfigure}%
        \begin{subfigure}[b]{0.33\linewidth}
    \includegraphics[width=1\linewidth]{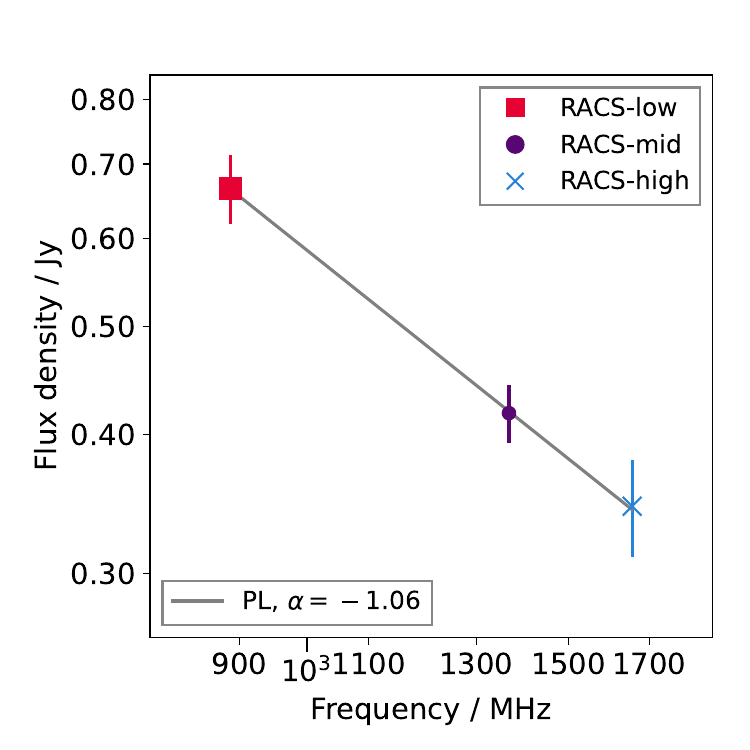}
    \caption{\label{fig:seds:j024} RACS-HIGH1~J024241$+$18564.}
    \end{subfigure}%
        \begin{subfigure}[b]{0.33\linewidth}
    \includegraphics[width=1\linewidth]{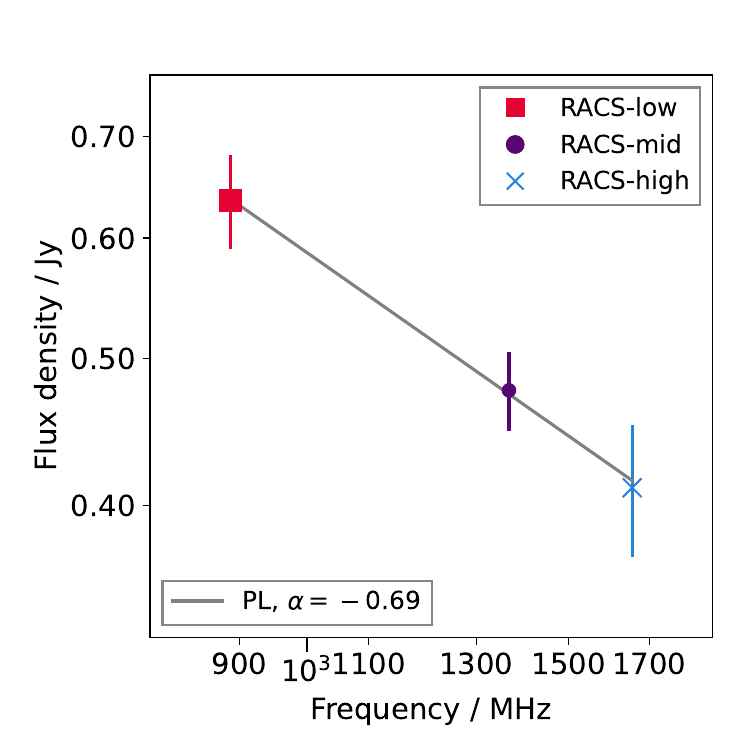}
    \caption{\label{fig:seds:j040} RACS-HIGH1~J040930$-$75334}
    \end{subfigure}\\%
        \begin{subfigure}[b]{0.33\linewidth}
    \includegraphics[width=1\linewidth]{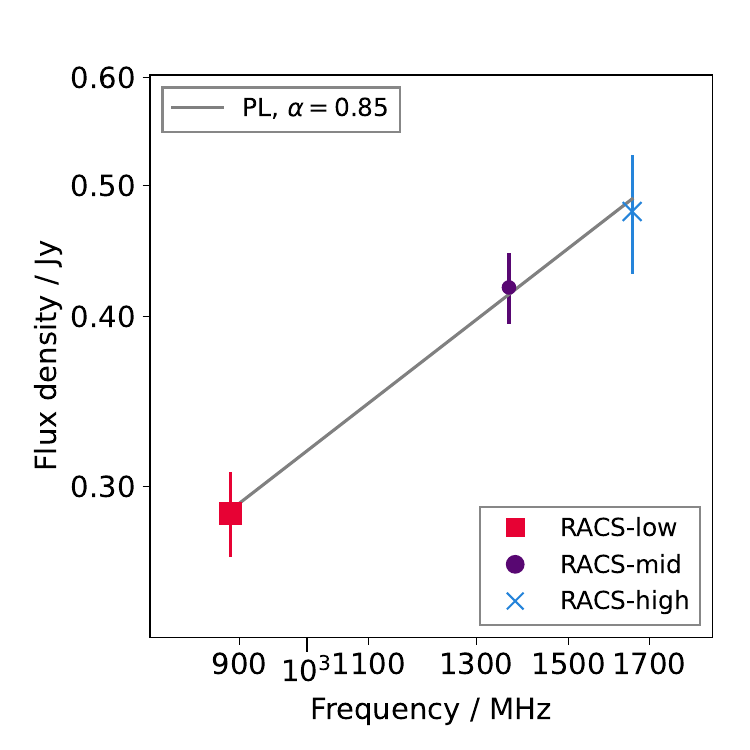}
    \caption{\label{fig:seds:j221} RACS-HIGH1~J221810$+$15203.}
    \end{subfigure}%
        \begin{subfigure}[b]{0.33\linewidth}
    \includegraphics[width=1\linewidth]{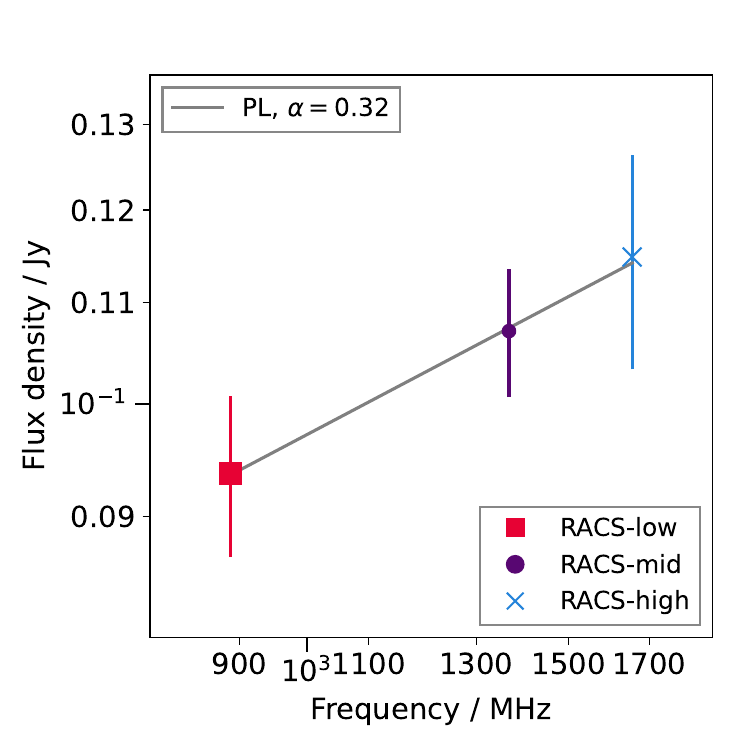}
    \caption{\label{fig:seds:j033} RACS-HIGH1~J033244$-$44555.}
    \end{subfigure}%
    \begin{subfigure}[b]{0.33\linewidth}
    \includegraphics[width=1\linewidth]{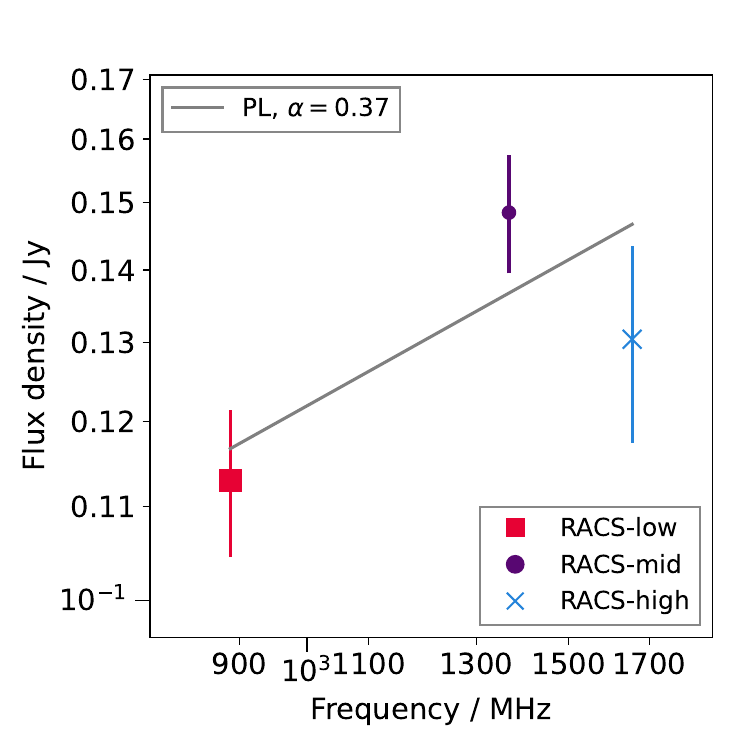}
    \caption{\label{fig:seds:j090} RACS-HIGH1~J090400$-$81514.}
    \end{subfigure}\\%
    \caption{\label{fig:seds} Example SEDs for cross-matched RACS sources. A power law (PL, with spectral index $\alpha$, solid grey line) is shown for each source.}
\end{figure*}

RACS epochs have provided useful measurements for wideband spectral modelling \citep[e.g.][]{Kerrison2024} in conjunction with other radio data, and with the currently released epochs, we have three flux density measurements for many sources---from RACS-low (887.5\,MHz), RACS-mid (\corrs{1\,367.5 MHz}), and RACS-high (\corrs{1\,655.5\,MHz}). By construction this frequency range provides good coverage across the range accessible to ASKAP so needs to be suitable to model source spectra for ASKAP calibration purposes. While we are not attempting to model all sources in this present work, in this section we assess the prospects of the three complete RACS epochs for spectral modelling. Future work will also incorporate RACS-low3 at 943.5\,MHz, for which a catalogue is currently being prepared to be used in upcoming linear polarization work associated with the SPICE-RACS project (Thomson et al., in preparation). 

\subsection{Source cross-matching and modelling}

We begin by again considering only unresolved sources in all three catalogues. We note that both in terms of morphological complexity and differences in RACS epoch angular scale sensitivity, extended sources will require additional consideration beyond the scope of this paper. For ease of this initial cross-matching, we also consider only isolated sources---i.e., those with no neighbours within 25\,arcsec, approximately the maximum PSF major FWHM in any of the catalogues in the overlap region ($\delta_\text{J2000} \lesssim +30\degr$). We then hierarchically match the catalogue source positions within 6\,arcsec, starting with RACS-high and decreasing in frequency. \corrs{For this initial assessment, we only consider sources with $>10\sigma_\text{rms}$ measurements across all three catalogues to avoid various low-$\sigma$ systematics (e.g.\ lower reliability and increased uncertainty in unresolved source determination). This fairly strict matching process results in 209\,104 sources.}  

Sources are then fit with a generic powerlaw model of the form \begin{equation}\label{eq:pl}
S = A\nu^\alpha \, \text{Jy} ,
\end{equation}
where $\alpha$ is the spectral index. Once we incorporate the RACS-low3 943.5-MHz measurements, we can also include modelling for curved spectra using generic curved power law model \citep[e.g.][and see \citealt{Lynch2021} and \citealt{Ross2024} for use of generic curved models in wideband spectral modelling]{db12}. For the purpose of this initial cross-matching, and with only three measurements currently, we opt to only consider the standard power law model. We use non-linear least-squares fitting through the Levenberg--Marquardt algorithm, incorporating uncertainties of 7\% for RACS-low (\citetalias{racs1}), 6\% for RACS-mid (\citetalias{racs-mid}), and 10\% for RACS-high (Section~\ref{sec:validation:flux}), added in quadrature to the measurement uncertainties reported in each catalogue, typically from the \texttt{PyBDSF} fitting procedure. 

\subsection{A brief analysis of the spectra}

Figure~\ref{fig:sed:hist} shows the distribution of the spectral indices ($\alpha$ from Equation~\ref{eq:pl}) from the cross-matched RACS sources. The mean $\alpha$ for all sources is $\bar{\alpha} = -0.77 \pm 0.48$ (with median $-0.85$). The mean/median $\alpha$ are fairly consistent with expected extragalactic populations \citep[e.g.][]{con92} and indeed consistent with our choice of $\alpha = -0.8$ earlier for brightness scale normalisation with respect to the NVSS. This is also similar to what we see in previous isolated RACS epoch comparisons with external surveys within this frequency range (\citetalias{racs2,racs-mid2}). Figure~\ref{fig:seds} shows six example sources and their spectra across the RACS band. Figures~\ref{fig:seds:j000}--\subref{fig:seds:j040} show a range of sources well-described by the power law model with $\alpha < 0$, and Figures~\ref{fig:seds:j221}--\subref{fig:seds:j090} show example sources with $\alpha > 0$. Figure~\ref{fig:seds:j090} is described as a flat-spectrum source in the Combined Radio All-Sky Targeted Eight GHz Survey \citep[CRATES;][]{crates}, cross-matched with the PMN~J0903$-$8151 \citep[considered a candidate flat-spectrum blazar by][]{DAbrusco2019}. 

We also compare the distribution against other widefield catalogues that have derived power law models for radio sources. Namely, the GLEAM-X DR2 catalogue \citep{Ross2024}, the NVSS-TGSS spectral index catalogue \citep{deGasperin2018}, and the SPECFIND V3.0 catalogue \citep[][but se also \citealt{Vollmer2005a,Vollmer2005b,Vollmer2010} for a description of the SPECFIND tool and early versions of the catalogue]{Stein2021}. The GLEAM-X DR2 catalogue is constructed with 20 narrowband measurements that are used for the spectral modelling---for \corrs{consistency and to avoid low-$\sigma$ systematics} we consider sources where all measurements are $>10\sigma_\text{rms}$. For the NVSS-TGSS catalogue, we also only consider sources with a measured two-point spectral index where both the NVSS and TGSS measurements are $>10\sigma_\text{rms}$. We do not have an equivalent cut for the SPECFIND V3.0 catalogue as that information is not readily available, though note that SPECFIND V3.0 incorporates the NVSS and TGSS in its spectral modelling, so there is some overlap with with the NVSS-TGSS spectral index catalogue.   

The distributions of $\alpha$ for these catalogues (with cuts) are shown on Figure~\ref{fig:sed:hist} alongside the RACS distribution. The mean $\alpha$ for the $10\sigma_\text{rms}$ GLEAM-X DR2 catalogue is $-0.84 \pm 0.20$ (with median $-0.84$, for 100\,598 GLEAM-X DR2 sources), and $ -0.75 \pm 0.22$ (with median $-0.76$ for the 325\,991 sources). While the distributions are broadly consistent, we note that the RACS source models show a larger standard deviation in the $\alpha$ distribution compared to GLEAM, GLEAM-X, and  NVSS-TGSS catalogues, likely a result of the sampling and fractional bandwidth of this work. GLEAM/GLEAM-X have both larger fractional bandwidth and finer sampling across the bandwidth, better informing the source spectra at the relevant frequencies. This enables better characterisation of non-powerlaw spectra (e.g. curvature) within the MWA band. {However, the higher frequencies probed by RACS is more suited to identifying peaked-spectrum sources (PSS) with a spectral peak $\gtrsim 1$\,GHz. Such sources may be too absorbed and/or faint at lower frequencies for GLEAM/GLEAM-X detection. An `inverted' positive spectral index are believed to be an identifier of compact young AGN, which are vital for understanding the evolution and life-cycle of large scale AGN, but require sufficient sampling below the spectral turnover for detailed spectral characterisation. Conversely, \corrs{f}or the NVSS-TGSS case, the larger difference in frequency between the two surveys provides reduces uncertainty in the two-point spectral indices. The additional 943.5-MHz measurement from RACS-low3 will improve this slightly.}

{It is worth noting, spectral index estimation using the NVSS-TGSS catalogue is particularly vulnerable to incorrectly estimating the spectral index due to long term ($\sim$decades) variability due to the large temporal gap between the NVSS and TGSS ($\sim$ two decades). The rapid survey speed of ASKAP ensures the time between RACS epochs remains relatively small, with all observations for RACS taken within five years. Consequently, the spectral modelling of this work has a reduced risk of incorrectly estimating the spectral index of sources due to any long-term variability. Furthermore, the timescales of observations for both RACS (2019--2024) and GLEAM-X (2018--2020), provides a broad spectral range (72--\corrs{1\,656}\,MHz) with complimentary temporal coverage. This makes a powerful combination for detecting and classifying PSS, which are known to show variability on timescales of months \citep{2022MNRAS.512.5358R} to decades \citep{2020ApJ...905...74N}, including losing their PSS classification entirely \citep{torniainen2005long}. A comprehensive analysis of PSS identified by combining GLEAM-X and RACS is beyond the scope of this work but will be described in Ross et al. (in preparation).}

\section{{Ongoing and future work with RACS}}\label{sec:future}

\corrs{The addition of RACS-high images into the RACS (and widefield radio survey) ecosystem provides the overall highest angular resolution survey covering the entire Southern Hemisphere at this sensitivity. While the RACS-high catalogue and full-sensitivity images are close in angular resolution to the equivalent data products from RACS-mid, the single SBID images from RACS-high have the highest angular resolution available from RACS, reaching $6\farcs5 \times 5\farcs5$. These images can be accessed through CASDA, with a cutout service available through the CASDA web interface or through the CASDA \texttt{astroquery} \footnote{\url{https://astroquery.readthedocs.io/en/latest/casda/casda.html}.} module. High angular resolution is an important property for radio images to allow modelling of source morphologies and de-blending from neighbouring, often unassociated, radio sources. This particularly benefits identification and cross-matching of radio sources with sources at other wavelengths \citep[e.g.][]{Broderick2024}, with ongoing projects already making use of the available RACS-high data \citep{Anumarlapudi2024,Driessen2024}.}

With the completion of the first epoch of RACS-high, work has begun in processing the archived second epoch of RACS-low (RACS-low2). These data at 888\,MHz mirror the observations from the first epoch of RACS-low, but with the \corrs{inclusion of} improvements to observation scheduling to constrain observations to $\pm 1$ hour angle off the meridian, peeling of bright, off-axis sources, and other telescope improvements in \corrs{the} intervening years. These data will be deposited onto CASDA once finished, and a catalogue will be made available in the same manner as \corrs{RACS-high shortly thereafter.} In addition to RACS-low2, the third epoch of RACS-low (RACS-low3) was also observed between 2023-12-20 \corrs{and} 2024-04-06 and was processed using the standard \texttt{ASKAPsoft} semi-automated pipeline used for other ASKAP projects. The initial images and calibrated visibilities were made available on CASDA along with validation reports for individual observations. \corrs{W}e have also constructed an all-sky catalogue from these RACS-low3 data following the process described in this work, which will be described in upcoming work (Thomson et al., in preparation). 

RACS-low3 differs from the previous low-band epochs in two key ways: the central frequency is 944\,MHz and the PAF footprint is more compact, matching RACS-mid and RACS-high. RACS-low3 was primarily observed to provide low-frequency observations to combine with RACS-mid and RACS-high within the context of full-band synthesis. Matching PAF footprints allows individual beams with identical pointing directions to be combined in a straightforward way. Combining all three bands will provide higher-sensitivity Stokes $I$ continuum images, and more importantly will also allow better spectral modelling for both total intensity and linear polarization. The larger bandwidth directly benefits the SPICE-RACS project. First, the larger bandwidth improves Faraday depth resolution for RM synthesis~\citep{bd05} (RACS-low3: $\sim60$\,rad\,m$^{-2}$, RACS-low3+mid+high: $\sim30$\,rad\,m$^{-2}$). Critically, the inclusion of higher frequencies significantly boosts the sensitivity to `Faraday thick' emission~\citep[the Faraday `max-scale',][]{bd05} which would otherwise be depolarised (RACS-low3: $\sim40$\,rad\,m$^{-2}$, RACS-low3+mid+high: $\sim110$\,rad\,m$^{-2}$). The shift in frequency to 944\,MHz was also chosen for other technical reasons, related to overall improved data quality in this part of the low band. A secondary benefit of the choice in frequency was to match the observational setup of the EMU and POSSUM projects, allowing easier scheduling as the observations could then share holography-derived primary beam models and calibrator scans.

\section{Summary}

We have described the next major data release for the Rapid ASKAP Continuum survey at \corrs{1\,655.5}\,MHz, entitled `RACS-high'. This additional epoch of RACS has been predominantly processed using the standard \texttt{ASKAPsoft} pipeline, including calibration, flagging, and imaging of Stokes $I$ and $V$ data products. As with RACS-mid, processing also involves a bespoke peeling pipeline used for a selection of observations with significant contamination from off-axis bright sources. These pipeline data products are made available through the CSIRO ASKAP Science Data Archive (CASDA) along with most other public ASKAP data products. 

As with RACS-low and RACS-mid, we produce a set of full-sensitivity mosaic images for each pointing. These images are used to create source lists that are merged into an all-sky catalogue. The images have a median resolution of $11\farcs8 \times 8\farcs1$ (Section~\ref{sec:resolution}) and a median rms noise of $195_{-32}^{+43}$\,\textmu Jy\,PSF$^{-1}$ (Section~\ref{sec:tile_sensitivity}). The catalogue has 2\,677\,509 sources (2\,461\,557 sources with $|b|>5\degr$) and covers the sky up to $+48\degr$ declination. We find that the reliability of the catalogue improves as the resolution degrades, similar to RACS-mid (Section~\ref{sec:reliability}). The catalogue (and full-sensitivity images) features a brightness scale uncertainty of 10\% (Section~\ref{sec:validation:flux}) and astrometric accuracy of $\approx 1$\,arcsec (Section~\ref{sec:astrometry}).  These full-sensitivity images and the all-sky catalogue are available through CASDA and are generally the recommended data product to use unless time-specific information is required.

We cross-matched the existing three RACS catalogues (RACS-low, RACS-mid, and RACS-high) to investigate the overall spectral features present in RACS sources so far (Section~\ref{sec:spectra}. We find an overall median spectral index, $\alpha=-0.85\pm0.48$, broadly consistent with other widefield and wideband radio catalogues with $\alpha$ measurement, though note that the three measurements for RACS limits precision of the spectral models. RACS has two additional epochs currently observed: RACS-low2 at 887.5\,MHz and RACS-low3 at 943.5\,MHz. RACS-low2 is currently being processed and will be released in a similar fashion to RACS-high. RACS-low3 has already been processed through the semi-automated \texttt{ASKAPsoft} pipeline soon after it was observed, with initial imaging data products already publicly available on CASDA. These additional epochs, particularly RACS-low3, will help improve the spectral modelling as we build a global sky model for ASKAP.

\section{Data availability}

As with other RACS data releases, we make available calibrated visibility datasets in the form of a \texttt{MeasurementSet} for each PAF beam, Stokes $I$ and $V$ images for each field, including CLEAN component model and residual images, and separate zeroth and first-order Taylor term images. Additional metadata related to the \texttt{ASKAPsoft} pipeline processing, including calibration data, primary beam models, scripts, and log files, are also available online for each observation. All of this is available through CASDA (\url{https://data.csiro.au/domain/casda}) under project code AS110. For users interested in only RACS-high data products, setting the frequency range to only cover \corrs{1\,655.5}\,MHz will limit to these observations. Image and visibility data are available at this persistent link: \begin{enumerate}
     \item[] \url{http://hdl.handle.net/102.100.100/374842?index=1}
\end{enumerate}
Individual pipeline-output source lists for each observation are available at: \begin{enumerate}
    \item[] \url{http://hdl.handle.net/102.100.100/374841?index=1}
\end{enumerate}

We also release a set of `full-sensitivity' mosaic images as described in Section~\ref{sec:mosaic}. We encourage general users to consider using these images, as they do not feature lowered sensitivity at the image edges, and they have also undergone brightness scaling as described in Section~\ref{sec:norm}. We also make available the radio source catalogue (and associated Gaussian component catalogue) constructed from the full-sensitivity images. The table column details are provided in Appendix~\ref{app:columns}. These data are available at: \begin{enumerate}
    \item[] \url{https://doi.org/10.25919/g3jd-av02}
\end{enumerate}

For source-finding and mosaicking for RACS-high, we use simple \texttt{python} pipelines contained in the \texttt{RACSsf} and \texttt{RACSmos} packages. These are available at \url{https://gitlab.com/Sunmish/racssf} and \url{https://gitlab.com/Sunmish/racsmos}, respectively. 

\begin{acknowledgement}
This scientific work uses data obtained from Inyarrimanha Ilgari Bundara, the CSIRO Murchison Radio-astronomy Observatory. We acknowledge the Wajarri Yamaji People as the Traditional Owners and native title holders of the Observatory site. CSIRO’s ASKAP radio telescope is part of the Australia Telescope National Facility (\url{https://ror.org/05qajvd42}). Operation of ASKAP is funded by the Australian Government with support from the National Collaborative Research Infrastructure Strategy. ASKAP uses the resources of the Pawsey Supercomputing Research Centre. Establishment of ASKAP, Inyarrimanha Ilgari Bundara, the CSIRO Murchison Radio-astronomy Observatory and the Pawsey Supercomputing Research Centre are initiatives of the Australian Government, with support from the Government of Western Australia and the Science and Industry Endowment Fund. 

This research has made use of the VizieR catalogue access tool, CDS, Strasbourg, France (DOI: 10.26093/cds/vizier). The original description of the VizieR service was published in \citet{vizier}. We used a range of \texttt{python} software packages during this work and the production of this manuscript, including \texttt{astropy} \citep{astropy:2018}, \texttt{matplotlib} \citep{Hunter2007}, \texttt{numpy} \citep{numpy,numpy2020}, \texttt{scipy} \citep{scipy}, and \texttt{cmasher} \citep{cmasher}.  Some of the results in this paper have been derived using the \texttt{healpy} \citep{healpy} and HEALPix package. We make use of \texttt{ds9} \citep{ds9} and \texttt{topcat} \citep{topcat} for visualisation, as well as the ``Aladin sky atlas'' developed at CDS, Strasbourg Observatory, France \citep{aladin1,aladin2} for obtaining catalogue data. For precision rounding in \LaTeX tables we used \texttt{to-precision}: \url{https://bitbucket.org/william_rusnack/to-precision/src/master/}. We make use of the \texttt{cubehelix} colourmap \citep{cubehelix}.

\end{acknowledgement}

%% file: appendices.tex
\renewcommand{\thesubfigure}{(\roman{subfigure})}
\renewcommand\thefigure{A\arabic{figure}}   
\renewcommand\thetable{B\arabic{table}}
\setcounter{figure}{0}
\setcounter{table}{0}

\section{Stokes V images}\label{app:stokesv}

\begin{figure}[t]
    \centering
    \includegraphics[width=1\linewidth]{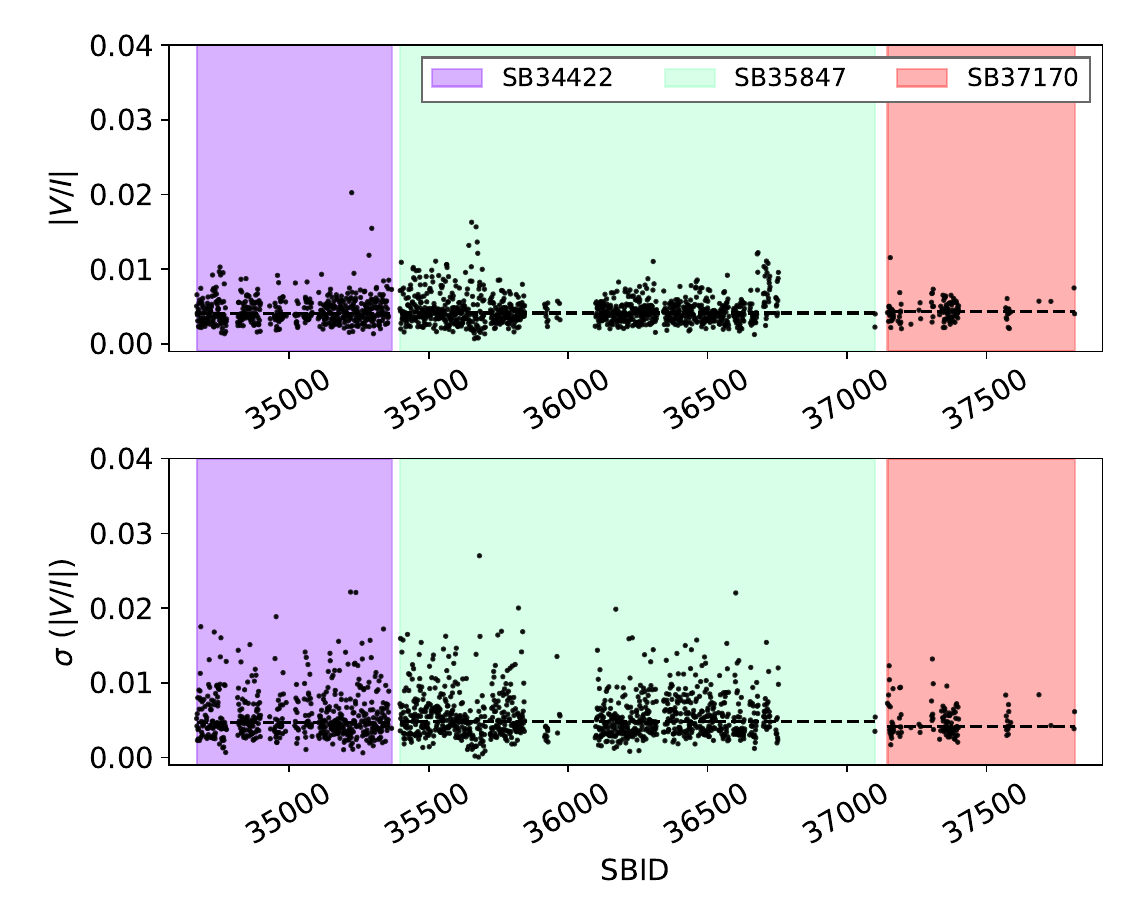}
    \caption{\label{fig:stokesv:sbid} \emph{Top.} The median leakage per observation recorded as the fractional circular polarization $|V/I|$ as a function of SBID for sources $>500\sigma_\text{rms,I}$. \emph{Bottom.} The standard deviation per obseravation as a function of SBID. The shaded regions show the time periods covered by the three separate primary beam models.}
\end{figure}

\begin{figure}[t]
    \centering
    \includegraphics[width=1\linewidth]{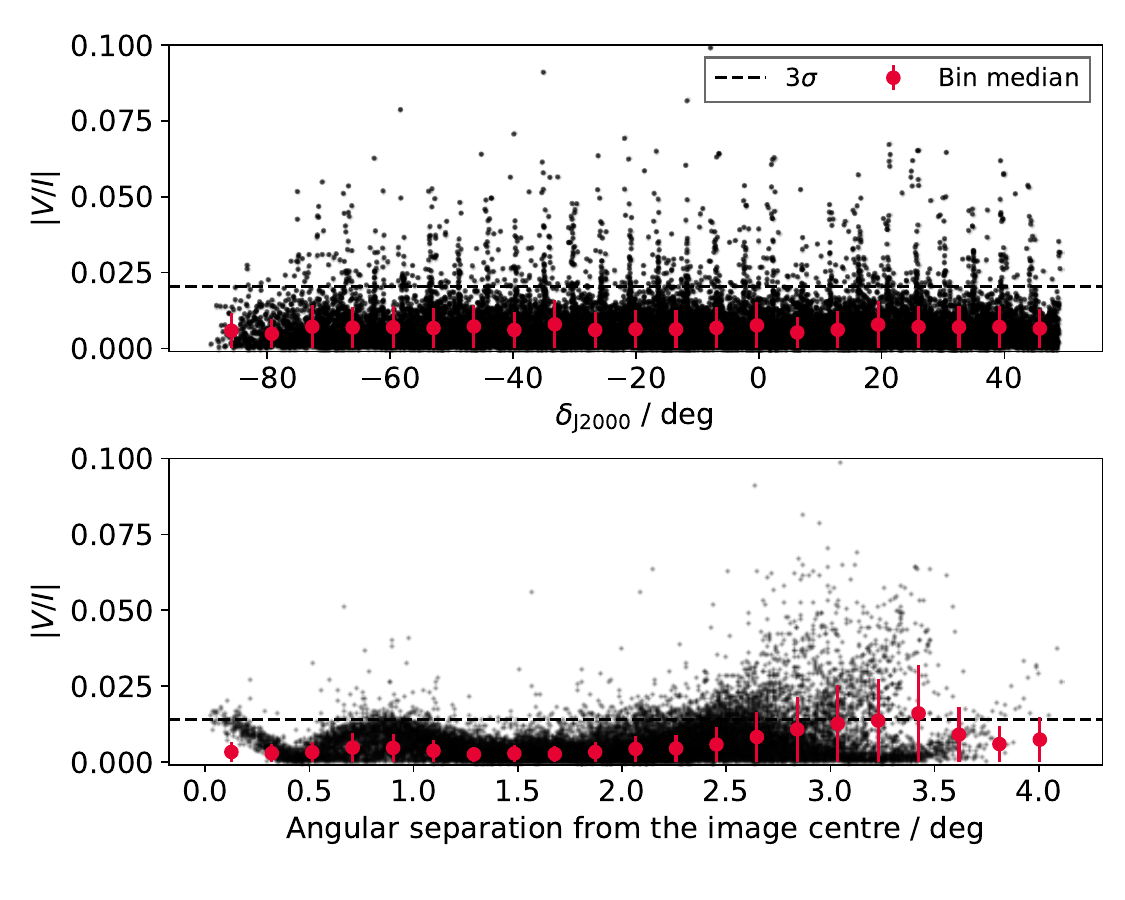}
    \caption{\label{fig:stokesv:coords} The fractional circular polarization for sources $>500\sigma_\text{rms,I}$ as a function of declination (\emph{top panel}) and angular separation from the image centre (\emph{bottom panel}). The red circles show median values within declination or angular separation bins with $\pm 1\sigma$ error bars. The horizontal black, dashed line indicates $3\sigma$.}
\end{figure}

Stokes $V$ images are produced independently of the Stokes $I$ images through a separate imaging step in the pipeline, with slightly different CLEAN settings (see section in 2.3.1 in \citetalias{racs-mid2}). The Stokes $V$ images are produced after application of an on-axis leakage correction derived from the bandpass calibrator PKS~B1934$-$638. However, widefield leakage correction is not performed due to the unavailability of holography-based widefield leakage models for Stokes V (see Section~\ref{sec:holography}). Note that the \texttt{ASKAPsoft} pipeline produces Stokes $V$ images with the IAU sign convention, where right-hand circularly polarised light is positive, and
left-hand circularly polarised light is negative. 

As part of validation processes we extract the peak pixel values in the Stokes V images at the locations of $>10\sigma_\text{rms,I}$ sources to assess the residual percentage leakage of Stokes I into V (as $V/I$ for what what we assume are sources with no circular polarization). Figure~\ref{fig:stokesv:sbid} shows the median $|V/I|$ per SBID as a function of SBID for sources with Stokes I signal-to-noise ratio (SNR) $>500$ (to avoid real circularly polarized sources). We also show the standard deviation of $|V/I|$ as a function of SBID. The overall mean and standard deviation is $|V/I| = (0.6 \pm 0.6)\,\%$ and $V/I = (-0.04 \pm 0.90)$\,\%, suggesting very minimal bias towards negative circular polarization. There is no substantial difference between the three major beam-forming periods recorded in Table~\ref{tab:holography}---the three periods highlighted on Figure~\ref{fig:stokesv:sbid} and labelled by their associated holography observation. 

Figure~\ref{fig:stokesv:coords} shows $|V/I|$ for all sources $>500\sigma_\text{rms,I}$ as a function of both declination (top panel) and angular separation from the image centre (bottom panel). Significant increases in the leakage can be seen at the image edges, and a slight ripple as is seen as a function of separation from the image centre. This ripple corresponds to the gaps between PAF beams (see Figure~\ref{fig:tile_sensitivity} where the same ripple is seen in sensitivity across the images, or in Figure~\ref{fig:tileflux:pre1} for the brightness scale). The implication is that the leakage in the individual PAF beams prior to mosaicking is significant and likely decreases due to averaging after mosaicking. The lack of widefield leakage corrections limits the reliability of the Stokes V images for sources. Within $\approx 2\degr$ of the image centres we can expect up to $\approx 2\,\%$ leakage, though this increases to $\approx 6$\,\% towards the image edges. The images are still suitable for some science cases \citep[e.g.~high-fractional polarization radio stars;][]{Pritchard2021} but care should be taken.

\section{Catalogue columns}\label{app:columns}

The columns for the RACS-high source and Gaussian component catalogues are a subset the columns in the RACS-mid catalogues (\citetalias{racs-mid2}, appendix A) and are reported in Table~\ref{tab:columns:source} and \ref{tab:columns:component}. \corrs{Note that surface brightness quantities are stored as Jy\,beam$^{-1}$ in the tables and are equivalent to the Jy\,PSF$^{-1}$ units used elsewhere in the text.}

\begin{table*}
    \centering
    \caption{\label{tab:columns:source} Columns in the RACS-high source catalogue.}
    \begin{adjustbox}{max width=\textwidth}
    \begin{tabular}{l l c l}\toprule
        Index & Name & Unit & Description \\\midrule
        0 & \texttt{Name} &  & Source name following IAU source name convention: \texttt{RACS-HIGH1 JHHMMSS.S$\pm$DDMMSS} \\
        1 & \texttt{Source\_ID} &  & Unique identifier for the source based on the \texttt{Field\_ID}. \\
        2 & \texttt{Field\_ID} & & Field name: \texttt{RACS\_HHMM$\pm$DD}. \\
        3 & \texttt{RA} & \corrs{deg} & J2000 right ascension of the source. \\
        4 & \texttt{Dec} & \corrs{deg} & J2000 declination of the source. \\
        5 & \texttt{Dec\_corr} & \corrs{deg} & J2000 declination, with declination-dependent offset correction (Section~\ref{sec:astrometry}). \\
        6 & \texttt{E\_RA} & \corrs{deg} & Uncertainty on \texttt{RA} from fitting the source position. \\
        7 & \texttt{E\_Dec} & \corrs{deg} & Uncertainty on \texttt{Dec} from fitting the source position. \\
        8 & \texttt{E\_Dec\_corr} & \corrs{deg} & Uncertainty on \texttt{Dec\_corr} from fitting the source position and declination offset model. \\
        9 & \texttt{Total\_flux} & mJy & Total flux density of the source. \\
        10 & \texttt{E\_Total\_flux\_PyBDSF} & mJy & Uncertainty in the total flux density from PyBDSF fitting. \\
        11 & \texttt{E\_Total\_flux} & mJy & Quadrature sum of the brightness scale and PyBDSF uncertainties for total flux density. \\
        12 & \texttt{Peak\_flux} & mJy\,beam$^{-1}$ & Peak flux density of the source. \\
        13 & \texttt{E\_Peak\_flux\_PyBDSF} & mJy\,beam$^{-1}$ & Uncertainty in the peak flux density from PyBDSF fitting. \\
        14 & \texttt{E\_Peak\_flux} & mJy\,beam$^{-1}$ &  Quadrature sum of the brightness scale and PyBDSF uncertainties for peak flux density. \\
        15 & \texttt{Maj\_axis} & arcsec & Size of the major axis of the source. \\
        16 & \texttt{Min\_axis} & arcsec & Size of the minor axis of the source. \\
        17 & \texttt{PA} & deg & Position angle of the source, east of north. \\
        18 & \texttt{E\_Maj\_axis} & arcsec & Uncertainty in the source major axis. \\
        19 & \texttt{E\_Min\_axis} & arcsec & Uncertainty in the source minor axis. \\
        20 & \texttt{E\_PA} & deg & Uncertainty in the source PA. \\
        21 & \texttt{DC\_Maj\_axis} & arcsec & Deconvolved size of the major axis of the source. \\
        22 & \texttt{DC\_Min\_axis} & arcsec & Deconvolved size of the minor axis of the source. \\
        23 & \texttt{DC\_PA} & deg & Deconvolved position angle of the source, east of north. \\
        24 & \texttt{E\_DC\_Maj\_axis} & arcsec & Uncertainty in the deconvolved source major axis. \\
        25 & \texttt{E\_DC\_Min\_axis} & arcsec & Uncertainty in the deconvolved source minor axis. \\
        26 & \texttt{E\_DC\_PA} & deg & Uncertainty in the deconvolved source PA. \\
        27 & \texttt{Noise} & mJy\,beam$^{-1}$ & Local estimate of the rms noise. \\
        28 & \texttt{Tile\_l} & deg & Direction cosine $l$ of the source with respect to the field centre. \\
        29 & \texttt{Tile\_m} & deg & Direction cosine $m$ of the source with respect to the field centre. \\
        30 & \texttt{Tile\_sep} & deg & Angular distance from the field centre. \\
        31 & \texttt{PSF\_Maj} & arcsec & FWHM of the PSF major axis of the field. \\
        32 & \texttt{PSF\_Min} & arcsec & FWHM of the PSF minor axis of the field. \\
        33 & \texttt{PSF\_PA} & deg & PA of the PSF of the field. \\
        34 & \texttt{S\_Code} & & Source structure classification provided by PyBDSF. \\
        35 & \texttt{N\_Gaussians} & & Number of Gaussian components comprising the source. \\
        36 & \texttt{Flag} & &  Source type flag. See Section~\ref{sec:reliability}. \\
         \bottomrule
    \end{tabular}
    \end{adjustbox}
\end{table*}

\begin{table*}
    \centering
    \begin{adjustbox}{max width=\textwidth}
    \caption{\label{tab:columns:component} Columns in the RACS-high component catalogue.}
    \begin{tabular}{l l c l}\toprule
        Index & Name & Unit & Description \\\midrule
        0 & \texttt{Gaussian\_ID} &  & Unique identifier for the component based on the \texttt{Field\_ID}. \\
        1 & \texttt{Source\_ID} &  & Unique identifier for the source the component belongs to (see Table~\ref{tab:columns:source}). \\
        2 & \texttt{Field\_ID} & & Field name: \texttt{RACS\_HHMM$\pm$DD}. \\
        3 & \texttt{RA} & \corrs{deg} & J2000 right ascension of the component. \\
        4 & \texttt{Dec} & \corrs{deg} & J2000 declination of the component. \\
        5 & \texttt{Dec\_corr} & \corrs{deg} & J2000 declination, with declination-dependent offset correction (Section~\ref{sec:astrometry}). \\
        6 & \texttt{E\_RA} & \corrs{deg} & Uncertainty on \texttt{RA} from fitting the component position. \\
        7 & \texttt{E\_Dec} & \corrs{deg} & Uncertainty on \texttt{DEC} from fitting the component position. \\
        8 & \texttt{E\_Dec\_corr} & \corrs{deg} & Uncertainty on \texttt{Dec\_corr} from fitting the component position and declination offset model. \\
        9 & \texttt{Total\_flux} & mJy & Total flux density of the component. \\
        10 & \texttt{E\_Total\_flux\_PyBDSF} & mJy & Uncertainty in the total flux density from PyBDSF fitting. \\
        11 & \texttt{E\_Total\_flux} & mJy & Quadrature sum of the brightness scale and PyBDSF uncertainties for total flux density. \\
        12 & \texttt{Peak\_flux} & mJy\,beam$^{-1}$ & Peak flux density of the component. \\
        13 & \texttt{E\_Peak\_flux\_PyBDSF} & mJy\,beam$^{-1}$ & Uncertainty in the peak flux density from PyBDSF fitting. \\
        14 & \texttt{E\_Peak\_flux} & mJy\,beam$^{-1}$ &  Quadrature sum of the brightness scale and PyBDSF uncertainties for peak flux density. \\
        15 & \texttt{Maj\_axis} & arcsec & Size of the major axis of the component. \\
        16 & \texttt{Min\_axis} & arcsec & Size of the minor axis of the component. \\
        17 & \texttt{PA} & deg & Position angle of the component, east of north. \\
        18 & \texttt{E\_Maj\_axis} & arcsec & Uncertainty in the component major axis. \\
        19 & \texttt{E\_Min\_axis} & arcsec & Uncertainty in the component minor axis. \\
        20 & \texttt{E\_PA} & deg & Uncertainty in the component PA. \\
        21 & \texttt{DC\_Maj\_axis} & arcsec & Deconvolved size of the major axis of the component. \\
        22 & \texttt{DC\_Min\_axis} & arcsec & Deconvolved size of the minor axis of the component. \\
        23 & \texttt{DC\_PA} & deg & Deconvolved position angle of the component, east of north. \\
        24 & \texttt{E\_DC\_Maj\_axis} & arcsec & Uncertainty in the deconvolved component major axis. \\
        25 & \texttt{E\_DC\_Min\_axis} & arcsec & Uncertainty in the deconvolved component minor axis. \\
        26 & \texttt{E\_DC\_PA} & deg & Uncertainty in the deconvolved component PA. \\
        27 & \texttt{Noise} & mJy\,beam$^{-1}$ & Local estimate of the rms noise. \\
        28 & \texttt{Tile\_l} & deg & Direction cosine $l$ of the component with respect to the field centre. \\
        29 & \texttt{Tile\_m} & deg & Direction cosine $m$ of the component with respect to the field centre. \\
        30 & \texttt{Tile\_sep} & deg & Angular distance from the field centre. \\
        31 & \texttt{PSF\_Maj} & arcsec & FWHM of the PSF major axis of the field. \\
        32 & \texttt{PSF\_Min} & arcsec & FWHM of the PSF minor axis of the field. \\
        33 & \texttt{PSF\_PA} & deg & PA of the PSF of the field. \\
        34 & \texttt{S\_Code} & & Host source structure classification provided by PyBDSF. \\
        35 & \texttt{Flag} & &  Host source type flag. See Section~\ref{sec:reliability}. \\
         \bottomrule
    \end{tabular}
    \end{adjustbox}
\end{table*}